\documentclass[journal]{IEEEtran}
\usepackage{comment}
\usepackage{verbatim}
\usepackage{amsmath}
\usepackage{booktabs}
\usepackage{graphicx} 
\usepackage{cite}
\usepackage{amsmath}
\usepackage{multirow}
\usepackage{amsmath,amssymb,amsfonts}
\usepackage{algorithmic}
\usepackage{graphicx}
\usepackage{mathtools}
\usepackage{textcomp}
\usepackage{array}
\usepackage[linesnumbered,ruled,vlined]{algorithm2e}
\usepackage{romannum}
\usepackage{xcolor}
\usepackage[colorlinks]{hyperref}
\usepackage{doi}
\usepackage{wrapfig}
\usepackage{mathtools}

\ifCLASSINFOpdf
\else
\fi
\begin{document}
\pagenumbering{arabic}
\title{Advancing Ultra-Reliable 6G: Transformer and Semantic Localization Empowered Robust Beamforming in Millimeter-Wave Communications}
%
%
%

\author{Avi Deb~Raha,~\IEEEmembership{Student Member,~IEEE,}
        Kitae~Kim,~\IEEEmembership{Student Member,~IEEE,}
        Apurba~Adhikary,~\IEEEmembership{Student Member,~IEEE,}
        Mrityunjoy~Gain,~\IEEEmembership{Student Member,~IEEE,}
        Zhu~Han,~\IEEEmembership{Fellow,~IEEE,}
        and~Choong Seon~Hong,~\IEEEmembership{Fellow,~IEEE}
\vspace{-10mm}
\thanks{Avi Deb Raha, Kitae Kim, Apurba Adhikary, Mrityunjoy Gain, and Choong Seon Hong are with the Department of Computer Science and Engineering, Kyung Hee University, Seoul 446-701, South Korea.  (e-mail: avi@khu.ac.kr; glideslope@khu.ac.kr; apurba@khu.ac.kr; gain@khu.ac.kr; cshong@khu.ac.kr).

Zhu Han is with the Electrical and Computer Engineering Department,
University of Houston, Houston, TX 77004 (email: hanzhu22@gmail.com).

Corresponding author: Choong Seon Hong (e-mail: cshong@khu.ac.kr).
}}

%
%

\markboth{}%
{Shell \MakeLowercase{\textit{et al.}}: Bare Demo of IEEEtran.cls for IEEE Journals}
%



\maketitle

\begin{abstract}
Advancements in 6G wireless technology have elevated the importance of beamforming, especially for attaining ultra-high data rates via millimeter-wave (mmWave) frequency deployment. Although promising, mmWave bands require substantial beam training to achieve precise beamforming.  While initial deep learning models that use RGB camera images demonstrated promise in reducing beam training overhead, their performance suffers due to sensitivity to lighting and environmental variations. Due to this sensitivity, Quality of Service (QoS) fluctuates, eventually affecting the stability and dependability of networks in dynamic environments. This emphasizes a critical need for robust solutions.  This paper proposes a robust beamforming technique to ensure consistent QoS under varying environmental conditions. An optimization problem has been formulated to maximize users' data rates. To solve the formulated NP-hard optimization problem, we decompose it into two subproblems: the semantic localization problem and the optimal beam selection problem. To solve the semantic localization problem, we propose a novel method that leverages the K-means clustering and YOLOv8 model. To solve the beam selection problem, we propose a novel lightweight hybrid architecture that combines a lightweight transformer with a CNN architecture through a weighted entropy mechanism. This hybrid architecture utilizes multimodal data sources to dynamically predict the optimal beams. A novel metric, Accuracy-Complexity Efficiency (ACE), has been proposed to quantify this. Six testing scenarios have been developed to evaluate the robustness of the proposed model. Finally, the simulation result demonstrates that the proposed model outperforms several state-of-the-art baselines regarding beam prediction accuracy, received power, and ACE in the developed test scenarios.
\end{abstract}
\IEEEpeerreviewmaketitle

\section{Introduction}
\IEEEPARstart{T}{he} forthcoming sixth-generation (6G) wireless communication networks represent a pivotal advancement in wireless technologies, aiming to deliver unparalleled throughput data services while ensuring mass connectivity with minimal power consumption. These networks are expected to provide a high degree of integration, develop more affordable devices, and enable an intelligent networking system to meet the growing demand for mobile devices and applications. Moreover, 6G networks are poised to undergo significant transformations by incorporating the benefits of extensive multiple input multiple output (MIMO) technologies and advanced frequency spectrums \cite{apurba_journal, TVT1}. This shift is not merely an incremental improvement but a substantial leap forward, designed to accommodate the growing needs of emerging applications such as intelligent transportation systems (ITS), the metaverse\cite{meng2024task}, augmented reality (AR), virtual reality (VR), and cloud gaming~\cite{EfficientDeep}. Maximizing the swift data transfer rates in higher frequency ranges, like millimeter-Wave (mmWave) and terahertz (THz), requires the deployment of extensive antenna arrays and the use of highly targeted, narrow beams \cite{charan2021vision2}. Adjusting these targeted beams, however, results in significant beam training overhead, which can severely compromise the system's quality of service (QoS), especially in high mobility scenarios~\cite{lider}.

Consequently, developing innovative solutions is essential for sustaining QoS in high-mobility contexts, such as vehicle-to-infrastructure (V2I) communications. The past decade has seen significant research efforts to tackle the challenges of mmWave beam training and channel estimation overhead~\cite{prev_01, prev_02, prev_03, prev_04}. Among these efforts, formulating adaptive beam codebooks has been a primary focus~\cite{prev_01, prev_02, prev_03, prev_04}. Although traditional approaches have made strides in decreasing the training overhead, the reductions are insufficient for systems utilizing large antenna arrays\cite{gourango_1}.  
Deep learning offers a promising solution to the challenges of beam training and channel estimation in mmWave communications, particularly within the framework of 6G networks \cite{semantic_beam, gourango_1, gourango_2}.
Several deep learning models have been proposed to reduce beam training overhead by utilizing diverse data types. 
In \cite{firstvision}, the authors used images from RGB cameras to predict optimal beam indexes, enhancing beamforming accuracy. However, this method is significantly influenced by environmental and lighting conditions variations and is limited to scenarios involving a single user. To address these limitations and enhance accuracy through diverse data sources, a novel approach combining vision and positional data was proposed in \cite{gourango_1} and \cite{gourango_2}. These studies applied a refined ResNet-50 model for precise beam index prediction. Specifically, \cite{gourango_1} integrated GPS data into the ResNet-50's fully connected (FC) layer, whereas \cite{gourango_2} employed a YOLOv3 model to identify the target vehicle within a group of vehicles.

Even though both methods utilize location data in conjunction with RGB images, environmental factors, and time-varying lighting conditions continue to pose challenges and hinder the QoS as the deep learning models extract features that may not be invariant to such changes. 
For instance, we fine-tune a ResNet-152 \cite{gainKCC} model by incorporating GPS features into the FC layer using scenario 3 data from the DeepSense 6G dataset~\cite{dataset}. This scenario encompasses low-light RGB image data captured from a base station (BS) camera, including images collected during the afternoon, evening, and nighttime. We employ low-light images (afternoon and evening) for training and nighttime images for testing.

\begin{figure}[t]
\centerline{\includegraphics[width=6cm]{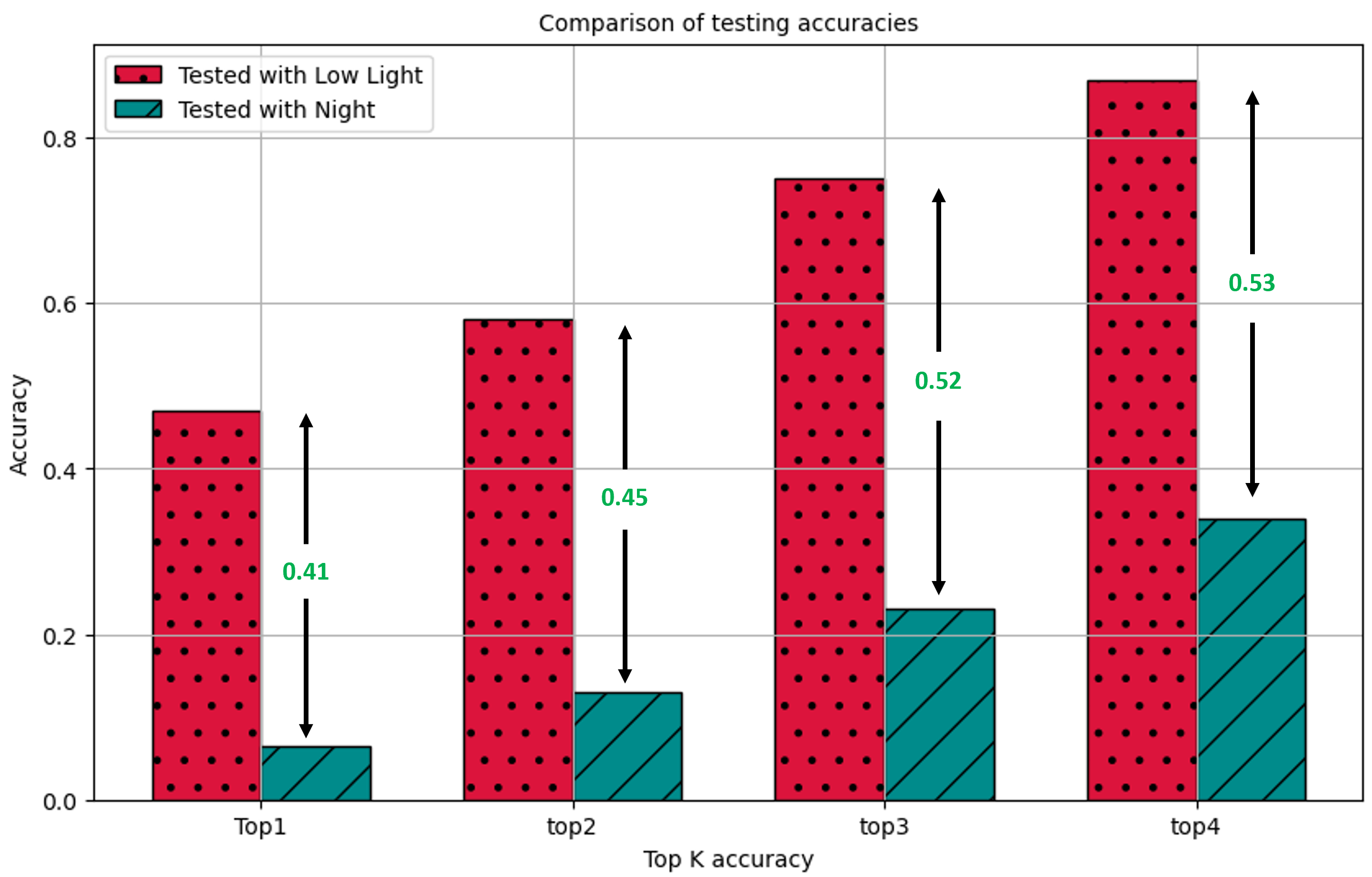}}
\vspace{-3mm}
\caption{Comparison of testing accuracy of low light images and night images using multimodal data.}
\label{daynight}
\vspace{-8mm}
\end{figure}

As illustrated in Fig. \ref{daynight}, the results indicate that even state-of-the-art (SOTA) deep learning models struggle to maintain performance despite integrating GPS features.
Furthermore, the computational complexity of models such as ResNet-50 and YOLOv3, used in \cite{gourango_1} and \cite{gourango_2}, is considerable. With parameter counts of approximately 25.6 million and 65 million, respectively, their extensive computational requirements make them less suitable for high-mobility applications, where efficiency and speed are essential. Consequently, using large models for beam prediction can also result in lower QoS, as the prediction delay may lead to sub-optimal beam usage.

Additionally, the environment is highly dynamic, and even minor changes in lighting conditions can significantly decrease the accuracy of optimal beam prediction, negatively impacting the QoS. Dynamic changes in the background, such as significant alterations in urban landscapes (e.g., removal or addition of large structures), can also affect performance.
Dynamic changes in the background, such as significant alterations in urban landscapes (e.g., removal or addition of large structures), can also affect performance. Therefore, we need a robust method to ensure high QoS under all conditions. 

Motivated by the visionary objectives of 6G technology and the advancements in artificial intelligence (AI), a new paradigm in communication, known as semantic communication (SemCom) \cite{farshbafan2023curriculum, Loc, raha2023, 9252948, bao2023mdvsc, semComMemory, avi}, has recently been introduced. This approach shifts the focus from the accurate retrieval of transmitted symbols to the precise understanding of the meaning conveyed by those symbols. Semantic communication aims to enhance communication efficiency and reliability, especially in environments where traditional communication systems face challenges due to dynamic changes or in scenarios requiring high levels of data interpretation and understanding.
Several key studies have laid the groundwork for semantic communication systems within this evolving field. In \cite{farshbafan2023curriculum}, the authors propose a novel goal-oriented semantic communication framework for next-generation wireless networks, enabling dynamic task execution between a speaker and a listener through a common semantic language of hierarchical belief sets. 
To handle multiple users with varying computing capacities, Loc et al. \cite{Loc} proposed a Swin Transformer-based dynamic semantic communication framework for multi-user scenario. In \cite{avi}, a semantic communication framework for high-altitude platform (HAP) assisted connected and autonomous vehicular (CAV) networks is proposed. 
The work of \cite{raha2023} introduces a generative AI driven semantic communication framework for transmitting sequential images or videos. 
In \cite{bao2023mdvsc}, common and individual features are extracted from video frames, for semantic video transmission. In  \cite{semComMemory}, the authors introduce memory into semantic communications to mimic human communications.

Most of the previous semantic communication studies \cite{farshbafan2023curriculum, Loc, raha2023, 9252948, bao2023mdvsc, semComMemory, avi} only focused on how data (i.e., image, text, audio) can be transmitted efficiently from transmitter to receiver. In the specific context of mmWave beamforming and the challenges posed by environmental changes, semantic communication offers a promising direction. In \cite{JsacBeam}, a semantic segmentation module has been used to extract semantic features, and then a ResNet model has been used to predict the optimal beams. However, the SOTA semantic segmentation models \cite{kirillov2023segment} are not robust to changes in the RGB images\cite{qiao2023robustness}. In another work \cite{semantic_beam}, the authors use semantic masks for predicting optimal beamforming vectors. In \cite{SegmentAnything}, the segment anything model (SAM) has been used to extract the semantic masks. However, both studies \cite{semantic_beam, SegmentAnything} focus exclusively on single-user scenarios, and the deep learning models used in these studies \cite{JsacBeam, SegmentAnything} have high parameter counts. Consequently, they are not energy efficient and may delay predicting optimal beams.
Additionally, none of the previous study has explored how to maintain reliable communication and high QoS when faced with different environmental challenges.

The preliminary version of our work was published in \cite{Avi_NOMS}, where we proposed a semantic-empowered robust beamforming approach. Although the semantic-based method is helpful in different environmental settings to maintain QoS, it suffers in extreme environmental conditions where the RGB camera can not capture more explicit images of the users. This limitation arises because the effectiveness of semantic extraction relies heavily on the quality and clarity of visual data. In conditions such as heavy fog, rain, or very low light, the RGB cameras' ability to provide detailed and distinguishable images is significantly reduced, impacting the precision of semantic masks and, consequently, the robustness of beamforming solutions.
The major differences between the current work and \cite{Avi_NOMS} are the addition of a hybrid deep learning approach leveraging transformer and semantic model, the introduction of a new metric, enhanced localization of the target user, and the evaluation of various lighting and extreme weather conditions.
The following is a summary of the contributions:
\begin{itemize}
\vspace{-1mm}
\item This paper proposes a robust beamforming technique utilizing CNN and transformer models for mmWave communication wherein a BS provides service to a mobile user selected from a pool of potential users. This method ensures consistent QoS under varying environmental conditions, making it ideal for high-frequency communication.
\item To maintain the QoS under different environmental conditions, we formulate an optimization problem. The primary objective of this optimization problem is to maximize the data rates received by the user.
\item To efficiently address the optimization problem, we divide it into two subproblems: the semantic localization problem and the optimal beam selection problem. To solve the semantic localization problem, we propose a novel method for localizing the target user from a set of users using a novel technique that combines K-means clustering with YOLOv8 for effective target identification.
\item To address the beam selection problem, we propose a novel hybrid architecture that combines two lightweight deep learning models. 
Then the decisions from both approaches are dynamically integrated using a weighted entropy mechanism.
\item To uphold performance, QoS and energy efficiency, we emphasize the need for models that are both lightweight and efficient in predicting optimal beams. 
Therefore, we introduce a novel metric Accuracy-Complexity Efficiency (ACE), which integrates model complexity into the evaluation framework, providing a more comprehensive assessment than mere accuracy comparisons.
\item Finally, the simulation results confirm the effectiveness of the proposed framework, achieving an increase in average received power ranging from 38.02\% to 42.10\% compared to baseline methods in dynamic environments, when considering the ground truth power values. Additionally, we achieved a 22.83\% increase in received power compared to our previous work \cite{Avi_NOMS}, attributed to the use of a transformer and CNN-empowered hybrid model that dynamically selects the optimal beam based on changing environments
\end{itemize}

The rest of the paper is laid out following the system model
in Section \ref{system_model}, the problem formulation in Section \ref{problem}.
Afterward, in Section \ref{solution}, the proposed algorithms for performing the solutions are described and simulation results are explained in Section \ref{results}. Finally, we represent the concluding remarks in Section \ref{conclusion}.
\vspace{-8mm}
\section{System Model} \label{system_model}
This study considers a scenario where a BS uses an RGB camera and GPS to capture user information. Then, using that information, the BS needs to select the optimal beams from the beam codebook for the target users. We utilize deep learning approaches to localize the target user and precisely identify the optimal beamforming vectors. 
In the following subsections, we first describe the system model in subsection \ref{system}, then illustrate the localization mechanism of the target vehicle in subsection \ref{localization_system}, and finally detail the dynamic beam selection model based on different environmental conditions using multimodal data sources in subsection \ref{beam_selection_system}.
\vspace{-5mm}
\subsection{System Model}\label{system}
\begin{figure}[t]
\centerline{\includegraphics[width=8cm]{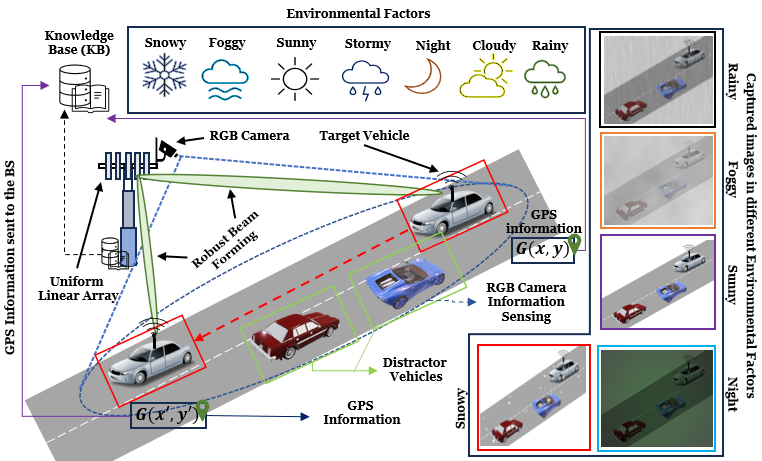}}
\vspace{-3mm}
\caption{System model for semantics empowered robust beamforming.}
\vspace{-3mm}
\label{System_Model}
\end{figure}
We consider a BS that is equipped with an RGB camera, a knowledge base (KB), a computing server, a GPS sensor, and an array of antennas ($A_K>>1$). 
The array of antennas is systematically organized in a uniform linear array (ULA) formation, facilitating optimal signal transmission and reception. 
The BS is assumed to operate with an analog architecture, incorporating a single RF chain, and employing a beamforming vector selected from a codebook denoted as $\textbf{X}$. The beamforming codebook can be represented as \cite{lider},
\vspace{-2mm}
\begin{equation}
\textbf{X}=[\mathbf{x}_1^L,\mathbf{x}_2^L, \dots,\mathbf{x}_i^L,\dots \mathbf{x}_N^L],\label{eq1}
\vspace{-2mm}
\end{equation}
where $N$ represents the total number of vectors in the set, and $L$ indicates the length of each beamforming vector. The $i^{th}$ beamforming vector from the codebook $\textbf{X}$ can be represented as \cite{EfficientDeep},
\vspace{-2mm}
\begin{equation}
\mathbf{x}_i^L = \frac{1}{\sqrt{K}} \left[ 1, e^{\left(j \frac{2\pi}{\mu} s \sin(\rho_{i})\right)}, \ldots, e^{j(K-1) \frac{2\pi}{\mu} s \sin(\rho_{i})} \right],
\label{eq2}
\end{equation}
where $K$ is the number of antennas elements, $s$ is the distance between adjacent antennas,
$\mu$ is the wavelength of the carrier, and the quantized azimuth angle of the beamforming vector is represented as $\rho_{i}$. 

The BS serves a mobile user, $v_i$, from a set of vehicles $\mathcal{V} = \{v_1, v_2, \ldots, v_i, \ldots, v_V\}$, with each vehicle equipped with a single antenna ($R_n=1$) and a GPS sensor, as depicted in Fig. \ref{System_Model}. In this setup, $v_i$ is considered the target vehicle, while the remainder of the set, $\mathcal{H} = \mathcal{V} \setminus {v_i} = \{v_1, v_2, \ldots, v_{i-1}, v_{i+1}, \ldots, v_V\}$, constitutes the distractor vehicles. 
The BS's objective is to precisely localize the target vehicle $v_i$ within the environment, effectively disregarding the distractor vehicles $\mathcal{H}$. 

To accurately capture and analyze the environment, the BS utilizes an RGB camera to take real-time images of the scene. 
We consider a narrowband block-fading channel model to describe the communication pathway from the BS to the user $v_i$. For the communication link between the BS and user $v_i$ on the $w^{th}$ subcarrier at time $t$, the channel vector $\mathbf{h}[t]$, composed of $Z$ distinct paths, is defined as follows \cite{lider}:
\vspace{-3mm}
\begin{equation}
\mathbf{h}_w[t] = \sum_{z=1}^{Z} \beta_z \mathbf{\alpha}(\rho_{z}^{az}, \rho_{z}^{el}),\label{eq3}
\vspace{-2mm}
\end{equation}
where the elevation angle and azimuth angle of path $z$ are represented by $\rho_{z}^{el}$ and $\rho_{z}^{az}$, respectively, and the complex gain is denoted as $\beta_z$. The array response vector, which functions based on both the elevation and azimuth angles, is expressed as $\mathbf{\alpha}(\rho_{z}^{az}, \rho_{z}^{el})$. 

For the received signal at the $w^{th}$ subcarrier and at time $t$ by user $v_i$ from the BS, is characterized as \cite{RGB_camera}:
\vspace{-2mm}
\begin{equation}
\mathbf{\tau}_w[t]=\textbf{h}^F_w[t]\textbf{x}^L_i[t]\delta_w[t]+y[t],\label{eq4}
\vspace{-2mm}
\end{equation}
where $\mathbf{\tau}_w[t] \in \mathbb{C}^{1 \times 1}$ represents the received signal at user $v_i$, $\textbf{x}^L_i \in \mathbb{C}^{N\times 1}$ denotes the beamforming vector, and $\textbf{h}^F_w[t] \in \mathbb{C}^{N\times 1}$ signifies the channel between the base BS and user $v_i$, $\delta_w[t] \in \mathbb{C}^{1\times 1}$ is the complex symbol transmitted from the BS to user $v_i$, and $y[t]$ is the additive white Gaussian noise originating from a complex zero-mean Gaussian distribution $\mathcal{N}(0, \sigma^2)$, where $\sigma$ indicates the standard deviation. The transmitted complex symbol $\delta \in \mathbb{C}^{1 \times 1}$ by the BS must adhere to the power constraint $\mathbb{E}[|\delta^2|] = P_{\delta}$, with $P_{\delta}$ representing the average symbol power. As we consider a single user, the average received data rate can be defined as\cite{semantic_beam},
\vspace{-2mm}
\begin{equation}
\textbf{W}[t]=\frac{1}{B}\sum_{b=1}^{B}\frac{P_\delta}{\sigma^2}|\textbf{h}^K_q[t]\mathbf{x}_i^L[t]|^2.\label{eq5}
\vspace{-2mm}
\end{equation}
$P_\delta$ is the average transmitted power for each complex symbol $\delta \in \mathbb{C}^{1 \times 1}$ and B is the number of subcarriers.
\vspace{-4mm}
\subsection{Semantic Localization Model}\label{localization_system}
For sensing and localizing the user, we consider an RGB camera installed in the BS. The images captured by the RGB camera can be expressed as \cite{avi},
\vspace{-2mm}
\begin{equation}
I:\vartheta_{t}\rightarrow[0,255]^c,\label{eq7}
\vspace{-2mm}
\end{equation}
where \(\vartheta_t= [[0;m-1]] \times [[0;n-1]]\) are pixels, and \(m\) and \(n\) stand for the total count of rows and columns. Additionally, for the RGB images \(c=3\) designates the total number of color channels. The RGB images can be also be considered as a time and environment-dependent function expressed as $I(t, \mathcal{E})$, where $t$ denotes time and $\mathcal{E}$ encapsulates a broad spectrum of environmental factors. These environmental factors include, but are not limited to, rain, storms, fog, and other atmospheric conditions. One of the primary challenges in using raw RGB images for beamforming is the variability caused by different weather conditions and lighting changes throughout the day. A model trained on daytime images may perform poorly at night and vice versa. This variability is further complicated by mixed conditions, such as a rainy afternoon or a foggy night, and the dynamic nature of urban environments with moving vehicles and architectural changes. Therefore, a semantic model is needed to consistently extract essential information from images regardless of environmental and temporal conditions, ensuring reliable beamforming in diverse settings.

The process for semantic localization of the user from RGB images can be outlined as follows:

1. \textbf{Vehicle Identification:}
The initial phase of the semantic localization process involves identifying all vehicles captured within the images. This step targets the vehicles of interest and accounts for potential distractors within the scene. The following expression can represent this process:
\vspace{-2mm}
\begin{equation}
\mathcal{J}: I(t, \mathcal{E}) \rightarrow \Omega_{t,\mathcal{E}}^\mathcal{V}; \quad \Omega_{t,\mathcal{E}}^\mathcal{V} = \mathcal{J}(I(t, \mathcal{E})). \label{eq8}
\vspace{-2mm}
\end{equation}
Here, $\Omega_{t,\mathcal{E}}^{\mathcal{V}}$ represents the identified vehicles, $\Omega_{t,\mathcal{E}}^{\mathcal{V}} = \{v_1, v_2, \ldots, v_i, \ldots, v_U\}$ within the image $I(t, \mathcal{E})$, effectively distinguishing these vehicles from other entities or background elements. The identification of vehicles  $\Omega_{t,\mathcal{E}}^{\mathcal{V}}$ is essential to ensure the accuracy and focus of subsequent analyses. For further processing and localize the target vehicle $v_i$, with each vehicle $v_j \in \Omega_{t,\mathcal{E}}^{\mathcal{V}}$ we retained the bounding box information $\xi_{v_j}^{bb}$ for vehicle $v_j$. Therefore,  (\ref{eq8}) can be represent as,
\vspace{-3mm}
\begin{align}
\mathcal{J}: I(t, \mathcal{E}) &\rightarrow \Omega_{t,\mathcal{E}}^\mathcal{V}; \nonumber \\
\Omega_{t,\mathcal{E}}^\mathcal{V} &= \{(v_1, \xi_{v_1}^{bb}), \ldots, (v_i, \xi_{v_i}^{bb}), \ldots, (v_U, \xi_{v_U}^{bb})\}
\label{eq9}
\end{align}
where $\xi_{v_i}^{bb} = \{(x_{min}^{v_i}, y_{min}^{v_i}, x_{max}^{v_i}, y_{max}^{v_i})\}$ denotes the bounding box of the $i^{th}$ vehicle, capturing the coordinates of the top-left and bottom-right corners of the rectangle encompassing the vehicle in the image space $I(t, \mathcal{E})$. The function $\mathcal{J}$ thus not only identifies the vehicles in the image but also keeps their bounding boxes.


   
   
2. \textbf{Comprehensive Environmental and Temporal Effect Mitigation}: Following the identification of vehicles, the next step addresses the challenge of environmental and temporal variability that can significantly impact the analysis. A sophisticated conversion of the identified vehicular data into a binary mask is performed to mitigate the effects of environmental and temporal changes. This binary mask only highlights essential features of the vehicles while minimizing the influence of varying conditions such as weather, lighting, and other unimportant objects (e.g., structures). This process can be represented as follows:
\vspace{-2mm}
\begin{equation}
\mathcal{H}: \Omega_{t,\mathcal{E}}^{\mathcal{V}} \to \alpha_{t,\mathcal{E}}^{\mathcal{V}} \in [0,1]; \quad \alpha_{t,\mathcal{E}}^{\mathcal{V}} = \mathcal{H}(\Omega_{t,\mathcal{E}}^{\mathcal{V}}).\label{eq2}
\vspace{-2mm}
\end{equation}
Here, $\alpha_{t,E}^{\mathcal{V}}$, is a binary mask that represents the vehicle contours, effectively discarding irrelevant background and environmental noise. This binary representation allows for a substantial reduction in the complexity of the images, focusing the analysis purely on vehicle attributes.

3. \textbf{Target Vehicle Detection}: To further enhance the precision of the data and localize the target vehicle \(v_i\) from the broader set of vehicles \(\mathcal{V}\), we employ an additional transformation on the binary mask. This transformation leverages the GPS location \(G^{v_i}(l_{\text{long}}^{v_i}, l_{\text{lati}}^{v_i})\) of the vehicle \(v_i\). This process can be represented as:
\vspace{-2mm}
\begin{align}
\small
\psi: \alpha_{t,E}^\mathcal{V}, \xi_{v_i}({G^{v_i}(\Phi_{\text{long}}^{v_i}, \Phi_{\text{lati}}^{v_i})}) &\rightarrow \alpha_{t,E}^{v_i} \in [0,1]; \nonumber \\
\begin{aligned}[t]
    \alpha_{t,E}^{v_i} &= \psi(\alpha_{t,E}^\mathcal{V}, \xi_{v_i}(G^{v_i}(\Phi_{\text{long}}^{v_i}, \Phi_{\text{lati}}^{v_i}))). 
\end{aligned}
\label{eq10}
\vspace{-1mm}
\end{align}
The binary mask \(\alpha_{t,\mathcal{E}}^\mathcal{V}\) is partitioned into equivalent segments \(\digamma = \{\chi_1,\chi_2,\ldots,\chi_i,\ldots,\chi_\digamma\}\). Utilizing GPS coordinates, the function \(\xi_{v_i}(\cdot)\) designates a specific segment, "\(\chi_i\)", from this partition. This process can be represented as,
\vspace{-2mm}
\begin{equation}
\chi_{i} = \xi_{v_i}(G^{v_i}(\Phi_{\text{long}}^{v_i}, \Phi_{\text{lati}}^{v_i})).
\vspace{-2mm}
\end{equation}
A distinct interval \((\chi_i-\Psi, \chi_i+\Psi)\) is then determined, encapsulating the target vehicle \(v_i\). Through the application of transformation \(\psi(\cdot)\), this interval is refined by excluding other vehicles not within the designated interval \((\chi_i-\Psi, \chi_i+\Psi)\) on the mask. When there is more than one bounding box in this interval, the selection is based on the area of the mask in the interval, which guarantees the single focus on the target vehicle \(v_i\).
This process optimization results in a mask \(\alpha_{t,E}^{v_i}\) that only contains the desired vehicle \(v_i\), thereby eliminating  distractor vehicles $v_j \in \mathcal{H}$. Focusing only on target vehicle $v_i$ improves beamforming accuracy and eliminates the need for large datasets that account for a range of environmental and temporal circumstances.
\vspace{-4mm}
\subsection{Beam Selection Model}\label{beam_selection_system}
Utilizing a hybrid architecture, our beam selection strategy incorporates two distinct models: one powered by semantic information $\alpha_{t, E}^{v_i}$ and the other by GPS data ${G^{v_i}(\Phi_{\text{long}}^{v_i}}, \Phi_{\text{lati}}^{v_i}))$. This approach is designed to sustain high-quality beamforming across diverse environmental conditions, ensuring that QoS is maintained even in adverse or extreme situations. The semantic model can perform better using detailed environmental cues for effective beam prediction. However, its reliance on camera data can cause problems in abysmal visibility or if the observation equipment is damaged, which makes gathering reliable semantic information challenging.

We incorporate a GPS data model to address these vulnerabilities and enhance the system's reliability. This integration provides a robust alternative data source unaffected by visual obstructions impacting the semantic model. Our system adapts to various environmental challenges by adopting this hybrid approach, ensuring efficient communication links and maintaining QoS despite external disturbances. The process of selecting beam using the semantic information $\alpha_{t,\mathcal{E}}^{v_i}$ can be represented as follows:
\vspace{-3mm}
\begin{equation}  \label{Optimal1}
\mathbf{x}_i^L = \mathbf{X}[s_1(\alpha_{t,\mathcal{E}}^{v_i})], 
\vspace{-3mm}
\end{equation}
where \(s_1(\alpha_{t,\mathcal{E}}^{v_i})\) denotes the model for predicting a beam index based on the semantic information encapsulated within \(\alpha_{t,\mathcal{E}}^{v_i}\), highlighting the significance of semantic insights in the beam selection process. The process of predicting optimal beam index using the location information $G^{v_i}(\Phi_{\text{long}}^{v_i}, \Phi_{\text{lati}}^{v_i})$ can be represented as follows:
\vspace{-2mm}
\begin{equation}  \label{Optimal2}
\mathbf{x}_i'^L = \mathbf{X}[s_2(G^{v_i}(\Phi_{\text{long}}^{v_i}, \Phi_{\text{lati}}^{v_i}))],
\vspace{-2mm}
\end{equation}
where \(s_2(\cdot)\) predicts the optimal beam index using GPS data, providing an alternative selection mechanism in instances where semantic data might not be dependable or available. 

Both \(s_1(\cdot)\) and \(s_2(\cdot)\) aim to accurately predict the optimal beamforming vector, which closely approximates the ground truth beamforming vector. Hence, it is important to optimize the loss function for both models. The loss function can be represented as \cite{attention_to_mPox, gain2024ccc}:
\vspace{-2mm}
\begin{equation}
\mathcal{L} = -\sum_{c=1}^{N} y_{o,\mathbf{x}_i^L} \log(p_{o,\mathbf{x}_i^L}), \label{loss}
\vspace{-2mm}
\end{equation}
where $\mathcal{L}$ denotes the loss for a single observation, $N$ is the number of beamforming vectors in codebook $\mathbf{X}$, $y_{o,\mathbf{x}_i^L}$ is a binary indicator of whether beamforming vector $\mathbf{x}_i^L$ is the optimal beam for observation $o$, and \(p_{o,\mathbf{x}_i^L}\) denotes the predicted probability that beamforming vector \(\mathbf{x}_i^L\) is the correct choice for observation \(o\). Both \(s_1(\cdot)\) and \(s_2(\cdot)\) will predict optimal beams based on the semantic data and location data, respectively. Then, their prediction will be used to select the final beam based on the following equations:
\vspace{-2mm}
\begin{equation}
\omega = 
\begin{cases}
1, & \text{if } \alpha_1 \leq \alpha_2,\\
0, & \text{otherwise},
\end{cases}
\vspace{-2mm}
\end{equation}

\begin{equation}
\mathbf{x}_i^{*L} = \omega \cdot \mathbf{x}_i^L + (1 - \omega) \cdot \mathbf{x}_i'^L. \label{Optimal}
\vspace{-2mm}
\end{equation}
Here, $\mathbf{x}_i^L$ is the selected beam from the $s_1(\cdot)$ model and $\mathbf{x}_i'^L$ is the selected beam from the second model $s_2(\cdot)$. This equation presents a weighted combination of the semantic localization-based and GPS data-based beam selection strategies, where \(\alpha_1\) and \(\alpha_2\) are the weights and based on the weights one beam will be chosen. By integrating these equations, the beam selection process underlines the
importance of utilizing multiple data sources for optimal operational efficiency, ensuring that the system can effectively navigate the challenges posed by different external conditions. Now, using the selected beam $\mathbf{x}_i^{*L} \in \mathbf{X}$ by (\ref{Optimal}), (\ref{eq5}) can be rewritten as: 
\vspace{-2mm}
\begin{equation}
\textbf{W}[t]=\frac{1}{B}\sum_{b=1}^{B}\frac{P_\delta}{\sigma^2}|\textbf{h}^K_q[t]\mathbf{x^*}_i^L[t]|^2.\label{eq_again_5}
\vspace{-2mm}
\end{equation}
\section{Problem Formulation}\label{problem}
In the system model under consideration, the primary goal is to enhance the average received data rates for the mobile user $v_i$, while maintaining an optimal QoS level across all environmental scenarios. This objective is achieved by employing both semantic localization information $\alpha_{t,\mathcal{E}}^{v_i}$ and GPS information $G^{v_i}(\Phi_{\text{long}}^{v_i}, \Phi_{\text{lati}}^{v_i})$. This leads to the formulation of an optimization problem in which the focus is to escalate the data rates, in accordance with (\ref{eq_again_5}), for the signal received by user \(v_i\). Accordingly, the optimization problem is expressed as:
\vspace{-3mm}
\begin{subequations}\label{Opt_1_4}
\vspace{-3mm}
    \renewcommand{\theequation}{\theparentequation\alph{equation}} 
	\begin{align}
	P1: \underset{{ \mathbf{x}_i^{L}, \mathbf{x}_i^{'L}, \chi_{v_i}, \omega,\Psi}}{\text{maximize}}
	&\;\sum_{t=1}^{T}\textbf{W}[t] \tag{18} \label{Opt_1_4:const} \\ 
	\text{subject to} \quad & \mathcal{C}1: \{\mathbf{x}_i^L, \mathbf{x}_i'^L\} \in \mathbf{X}, \label{Opt_1_4:const1} \\
	&\begin{aligned}[t]
	    \mathcal{C}2: \mathcal{T}^{ext}_{\psi^{v_i}_{t,\mathcal{E}}}\big(&\psi(\alpha_{t,\mathcal{E}}^\mathcal{V},\xi_{v_i}(G^{v_i}(\Phi_{long}^{v_i}, \\
	    & \Phi_{lati}^{v_i})))\big) \leq T_{max},
	\end{aligned} \label{Opt_1_4:const2} \\
	&\mathcal{C}3: \omega \in \{0, 1\} \label{Opt_1_4:const3} \\
    &\begin{aligned}[t]
        \mathcal{C}4: \chi_i \cap \chi_j &= \emptyset, \\
        & \forall i, j \in \{1, 2, \ldots, \digamma\}
    \end{aligned} \label{Opt_1_4:const4} \\
    &\mathcal{C}5: \Omega_{t,\mathcal{E}}^\mathcal{V} \in \mathcal{V},  \forall_{t,\mathcal{E}} \in I(t, \mathcal{E}) \label{Opt_1_4:const5} \\
    &\mathcal{C}6: \alpha_{t,\mathcal{E}}^{v_i} \in \{0,1\}, \forall_{t,\mathcal{E}} \in I(t, \mathcal{E})  \label{Opt_1_4:const6} \\
    & \mathcal{C}7: \Upsilon_{\min}, \Gamma_{\min} \leq l_{\text{{long}}}^{v_i}, l_{\text{{lati}}}^{v_i}  \nonumber \\ 
       & \hspace{5mm}\leq \Upsilon_{\max}, \Gamma_{\max}, \label{Opt_1_4:const7} \\
    &\mathcal{C}8: 1 \leq \chi_{v_i}, \Psi \leq X \label{Opt_1_4:const8}, \\
    &\mathcal{C}9: 1 \leq \chi_{v_i}-\Psi, \chi_{v_i}+\Psi \leq X \label{Opt_1_4:const9}, \\
    &\mathcal{C}10: 0 \leq \left\|\xi_{v}^{bb}\right\| \leq M, \label{Opt_1_4:const10} \\
    &\mathcal{C}11: 0 \leq \alpha_1 , \alpha_2 \leq \mathcal{K}, \label{Opt_1_4:const11} \\
    &\mathcal{C}12: {\mathcal{L}}: 0 < p_{o,\mathbf{x}_i^L} \leq 1, \forall o, \mathbf{x}_i^L. \label{Opt_1_4:const12}
	\end{align}
\end{subequations}
\begin{figure*}[h]
\centering
\includegraphics[width=17cm]{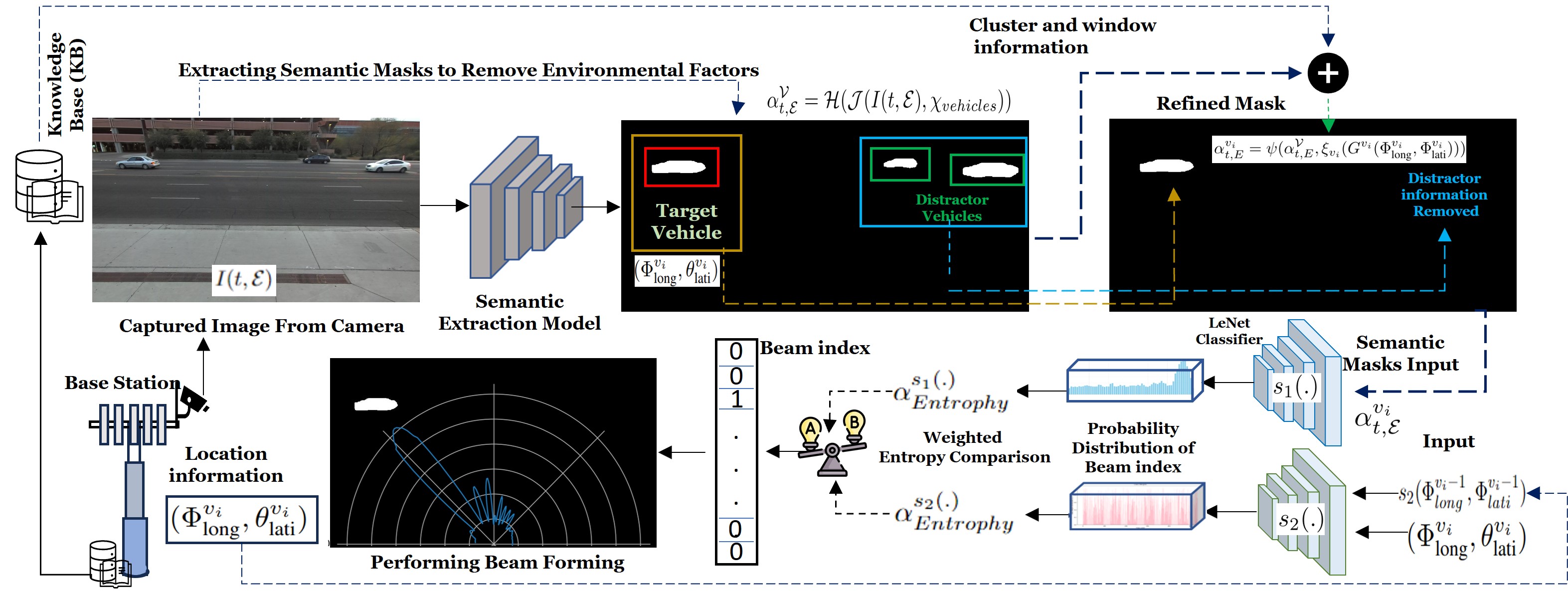}
\vspace{-3mm}
\caption{Step by step deep learning-based solution approach.}
\label{Solution approach}
\vspace{-4mm}
\end{figure*}
We consider five decision variables $\mathbf{x}_i^{L}$, $\mathbf{x}_i^{'L}$, $\chi_{v_i}$, $\omega$ and $\Psi$ in (\ref{Opt_1_4})  for maximizing the received data rate for every time step of user $v_i$. $\chi_{v_i}$ and $\Psi$ ensures an window over the mask $\alpha_{t,\mathcal{E}}^{\mathcal{V}}$ so that we can extract the semantic mask $\alpha_{t,\mathcal{E}}^{v_i}$ that contains only the information about target user $v_i$. Ensuring that the target vehicle $v_i$ is precisely localized within the range of $(\chi_{v_i}-\Psi)$ to $(\chi_{v_i}+\Psi)$ enables the semantic model $s_1(\cdot)$ to determine potentially effective beams, thus improving data rates. $\mathbf{x}_i^{L}$ and $\mathbf{x}_i^{'L}$ select the optimal beams for the user $v_i$ using (\ref{Optimal1}) and (\ref{Optimal2}) respectively. $\omega$ is a binary decision variable for choosing the final optimal beam $\mathbf{x}_i^{*L}$. Constraint $(\ref {Opt_1_4:const1})$ ensures that the beamforming vectors $\mathbf{x}_i^L$ and $\mathbf{x}_i'^L$ are predefined. Constraint $(\ref {Opt_1_4:const2})$ limits the time of the semantic localization process to the threshold $T_{max}$.
$(\ref {Opt_1_4:const3})$ ensures that only one of the beamforming vectors is selected based on the weights of the two models.
$(\ref {Opt_1_4:const4})$ guaranties the mutual exclusiveness of the division of the image plane.
$(\ref {Opt_1_4:const5})$ ensures the extraction vehicles should be from a predefined vehicle set $\mathcal{V}$. That helps the semantic extraction process to extract the required semantic information.  
$(\ref {Opt_1_4:const6})$ represents that $ \alpha_{t,\mathcal{E}}^{v_i}$ should be a binary mask.
Constraint $(\ref {Opt_1_4:const7})$ sets a lower limit and upper limit of the GPS position of the target vehicle $v_i$.
Constraints  $(\ref {Opt_1_4:const8})$ and $(\ref {Opt_1_4:const9})$ ensures that the window \( (\chi_{v_i}-\Psi, \chi_{v_i}+\Psi) \) remain on the image plane. Constraint  $(\ref {Opt_1_4:const10})$ ensures that the number of the detected vehicles from $I(t, \mathcal{E})$ is a finite number. Constraint  $(\ref {Opt_1_4:const11})$ sets an upper limit and lower limit on the weights $\alpha_1$ and $\alpha_2$. Constraint  $(\ref {Opt_1_4:const12})$ ensures that the predicted probabilities are within the valid range (0, 1), which is necessary for the computation of the cross entropy loss.

The formulated problem in $P1$ directs to a dynamic programming problem, which is employed for the stochastic environment as the vehicles are randomly distributed over time \cite{apurba, munir_np_hard}. Given this randomness, the model must accommodate a potentially expansive and unbounded number of states and transitions. This expansive state space and the complexity of the transitions are indicative of NP-hard problems.
The complexity of NP-hard problems is typically high, and it is hard to solve the problem in polynomial time \cite{munir_np_hard}. Additionally, the process of extracting semantic information relies on deep learning techniques. Therefore, to devise the solution to the formulated problem $P1$, we decomposed the problem into two sub-problems. The first sub-problem deals with the semantic localization $\alpha_{t,\mathcal{E}}^{v_i}$ from image $I(t, \mathcal{E})$ considering the constraints (\ref{Opt_1_4:const2}), (\ref{Opt_1_4:const4}), (\ref{Opt_1_4:const5}), (\ref{Opt_1_4:const6}), (\ref{Opt_1_4:const7}), (\ref {Opt_1_4:const8}), (\ref {Opt_1_4:const9}) and (\ref{Opt_1_4:const10}) for user semantic localization using decision variable $\chi_{v_i}$ and $\Psi$. On
the contrary, the second sub-problem deals with selecting the optimal beams for user $v_i$ from semantic mask $\alpha_{t,\mathcal{E}}^{v_i}$ considering the constraints (\ref{Opt_1_4:const1}), (\ref{Opt_1_4:const3}), (\ref{Opt_1_4:const7}), (\ref {Opt_1_4:const11}) and (\ref {Opt_1_4:const12}) using decision variables $\mathbf{x}_i^{L}, \mathbf{x}_i^{'L}$ and $\omega$. By selecting the optimal beam $\mathbf{x}_i^{*L}$ using the decision variables  $\mathbf{x}_i^{L}, \mathbf{x}_i^{'L}$ and $\omega$, the data rates can be maximized. This decomposition not only eases the computational burden but also allows for specialized algorithms to be applied to each sub-problem, potentially leading to more efficient and accurate solutions.
\vspace{-8mm}
\section{Solution approach}\label{solution}
Generally, for mmWave systems to attain an adequate SNR, it's essential to utilize large antenna arrays and deploy highly focused, narrow beams. To identify the optimal beams, users need to transmit an initial pilot signal upstream to the BS. Then, the uplink pilot signals are used to train the $N$ beams. However, the training cost for this process is very high,\cite{JSAC, gourango_1, semantic_beam, gourango_2}. This complexity of the pilot training or large beam scans highlights the need for efficient and effective solutions for identifying optimal beams, enhancing the performance of wireless communication applications with high mobility. Therefore, deep learning techniques emerge as a promising solution, leveraging past observations and supplementary external data to rapidly predict mmWave beams from the predefined beam codebook $\mathbf{X}$, thus guaranteeing reliability in V2I communications \cite{JSAC, gourango_1, semantic_beam, gourango_2}. To leverage deep learning for beamforming, a dataset $\mathcal{D}$ is required, which is defined as $\mathcal{D} = \{d_1, d_2, \ldots, d_i, \ldots, d_D\}$. In this context, each dataset entry $d_i$ is structured as $d_i = \{I(t,\mathcal{E})_i, \mathbf{x}_i^{L*}\}$, where $I(t,\mathcal{E})_i$ denotes the $i^{th}$ RGB image within the dataset, and $\mathbf{x}_i^{L*}$ signifies the optimal beam or the ground truth beam associated with that specific image.

The performance of the deep learning model heavily depends on the dataset. Therefore, the data instances on the dataset $D$ require all possible environmental data (i.e., sunny, rainy, snowy), time-varying data (i.e., day, night, evening), as well as their combinations data (i.e., sunny day, rainy day, rainy night) for performing robust beamforming in every condition. However, this is an impractical and impossible task. Therefore, we need a semantic extraction from data instances to neutralize the environmental and time-varying factors for every condition. The semantic localized data instances extracted from $\mathcal{D}$ can be represent by $d_i=\{\alpha_{t,\mathcal{E}}^{v_i}, \mathbf{x}_i^{L*}\}$. Then, the deep learning model is constructed to learn an estimation function $x_i^{*L}=f_{\theta}(\alpha_{t, \mathcal{E}}^{v_i})$. Here, the parameters of the deep learning model have been represented by $\theta$, and $x_i^{*L}$ represents an estimated beamforming vector. Formally it can be represented as \cite{SegmentAnything}, 
\vspace{-2mm}
\begin{subequations}\label{Opt_1_8}
\vspace{-2mm}
	\begin{align}
	f^*_{\theta^*}=\operatorname*{argmax}_{f_{\theta}}
	&\;\prod_{i = 1}^{D}\mathbb{P}(\mathbf{x}_i^{*L}=\mathbf{x}_i^{L*}|\alpha_{t,E}^{v_i})\tag{\ref{Opt_1_8}}.
	\end{align}
\end{subequations}
The objective is to find the best-fitting parameters, denoted as $\theta^*$, which maximize the likelihood of accurate prediction. The more precisely we can predict the optimal beam index, the more we can maximize the received power for the end-user, thus achieving the targeted outcome outlined in problem P1. Fig. \ref{Solution approach} shows the step by step solution approach. Next, we describe the semantic information extraction and beam selection process.
\vspace{-3mm}
\subsection{Semantic Localization Process}\label{AA}
The semantic localization process starts by detecting vehicles in real-time and converting their segments into binary masks for precise localization. These semantic masks, combined with GPS data, enable the identification of specific target vehicles through a K-means clustering algorithm. The details are described as follows:

1. \textbf{Vehicle Detection} and \textbf{Environmental Effect Removal}:
In this study, we use the pre-trained YOLOv8 model for vehicle detection due to its real-time detection capability and speed. To achieve rapid semantic localization of the user, we chose the streamlined YOLOv8n variant, which has 3.5 million parameters, balancing performance and efficiency. The single-pass architecture of YOLOv8n ensures effective and precise object identification, making it ideal for tasks requiring rapid responses. The pre-trained YOLOv8n model provides segments of vehicles, which enables the transformation of these segments into binary masks. This process can be represented as,
\vspace{-2mm}
\begin{equation}
\alpha_{t,\mathcal{E}}^\mathcal{V} = \mathcal{H}(\mathcal{J}(I(t, \mathcal{E}), \chi_{vehicles})).\label{eq14}
\vspace{-2mm}
\end{equation}
Here, $\mathcal{J}(\cdot)$ symbolizes the pretrained YOLOv8n model tasked with segmenting the input image $I(t, \mathcal{E})$ by identifying regions corresponding to the set of vehicle classes $\chi_{vehicles}$. The transformation function $\mathcal{H}(\cdot)$ is subsequently applied to these segments, adjusting the pixel values within the segmented areas to 1 (indicative of vehicle presence) and setting all other pixels to 0 (indicative of absence). This delineation process is essential for isolating vehicles from their surroundings, allowing for more precise and optimal beams.
The transformation mechanism can be detailed further as follows. For an input image $I(t, \mathcal{E})$, let $S = \mathcal{J}(I(t, \mathcal{E}), \chi_{vehicles})$ denote the segmentation result highlighting vehicle areas. The transformation function $\mathcal{H}(\cdot)$ converts the image into a binary mask $\alpha_{t,\mathcal{E}}^\mathcal{V}$, where the pixel values are defined by:
\vspace{-2mm}
\begin{equation}
\alpha_{t,\mathcal{E}}^\mathcal{V}(i,j) = 
\begin{cases} 
1 & \text{if } p_{i,j} \in S, \\
0 & \text{otherwise}, \label{eq20}
\end{cases}
\vspace{-2mm}
\end{equation}
where $p_{i,j}$ represents the pixel at position $(i, j)$ in the original image $I(t, \mathcal{E})$.

2. \textbf{Target Vehicle Detection}: 
The mask $\alpha_{t,\mathcal{E}}^\mathcal{V}$ can contain multiple vehicles \( \mathcal{V} \) from the set of classes \( \chi_{\text{vehicles}} \). To isolate the targeted vehicle \( v_i \), we utilized the GPS position and beam forming vector index $i_{x_i^{L*}}$ pair, \( (G^{v_i}(\Phi_{\text{long}}^{v_i}, \theta_{\text{lati}}^{v_i}), i_{x_i^{L*}}) \) of the target vehicles \( v_i \) from the training dataset. We divide the masked image \( \alpha_{t,E}^\mathcal{V} \) into a set of  \( \digamma = \{\chi_1,\chi_2,\ldots,\chi_i,\ldots,\chi_\digamma\} \) divisions as shown in Fig. \ref{beam_index} where each division has then been indexed to \( 1,\dots, \digamma \).
Then, we employ a K-means clustering algorithm \cite{K_means} on the GPS locations. Let \( \mathcal{P} = \{ p_1, p_2, \dots, p_N \} \) be the set of GPS locations where each \( p_i = (\Phi_{\text{long}}^{v_i}, \theta_{\text{lati}}^{v_i}) \). First, each location \( p_i \) is standardized as:

\begin{figure}[t]
\centerline{\includegraphics[width=6cm]{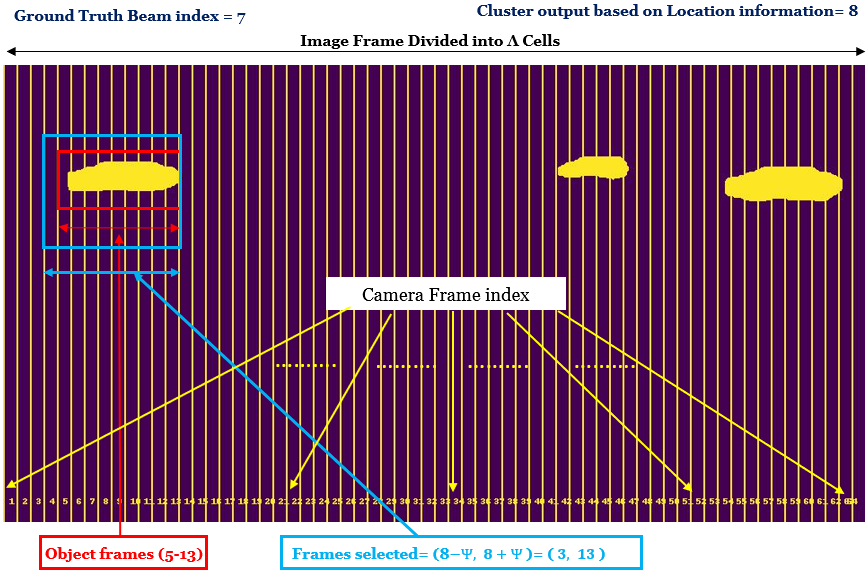}}
\vspace{-3mm}
\caption{Process of finding the target vehicle using K-means clustering.}
\vspace{-4mm}
\label{beam_index}
\end{figure}
\vspace{-2mm}
\begin{equation}
p_i' = \frac{p_i - \eta}{\omega},
\vspace{-1mm}
\end{equation}
where $\eta$ and $\omega$ represents the mean and standard deviation of the GPS locations $\mathcal{P}$ which can be calculated as follows:
\begin{equation}
\eta = \frac{\sum_{i=1}^{N} \Phi^{v_i}}{N},
\end{equation}
\begin{equation}
\omega = \sqrt{\frac{\sum_{i=1}^{N} (\Phi^{v_i} - \eta_{\Phi^{v_i}})^2}{N}}.
\end{equation}
The K-means objective function \( J \) aims to minimize the intra-cluster distances\cite{K_means}:
\vspace{-2mm}
\begin{equation}
J = \sum_{i=1}^{N} \min_{\mu_j \in C} \| p_i' - \mu_j \|^2,
\vspace{-1mm}
\end{equation}
where $\mu_j$ is a centroid and $C$ is the set of centroids. The algorithm assigns each $p_i'$ to the nearest cluster centroid \( \mu_k \):
\begin{equation}
S_k = \{ p_i' \mid \| p_i' - \mu_k \| \leq \| p_i' - \mu_j \| , \; \forall j, 1 \leq j \leq \digamma \}.
\end{equation}
The cluster centroids are then updated as:
\vspace{-2mm}
\begin{equation}
\mu_k = \frac{1}{|S_k|} \sum_{p_i' \in S_k} p_i'.
\vspace{-1mm}
\end{equation}
Finally, we map each cluster \( S_k \) to a division \( \chi_i \) based on the mode of the index values of $i_{\mathbf{x}_i^{L*}}$ within that cluster:
\vspace{-2mm}
\begin{equation}
\chi_{i}(S_k) = \arg\max_{x} \left| \{ p_i' \mid i_{\mathbf{x}_i^{L*}} = x, p_i' \in S_k \} \right|.
\vspace{-1mm}
\end{equation}
These updated centroids $\mu_k$ from the training dataset and their assigned division will be stored in the BS's KB $\mathcal{K}$. Whenever a vehicle $v_i$ enters the area of the BS, the BS will use the location of that vehicle to select a centroid $\mu_j$ based on the location $(G^{v_i}(\Phi_{\text{long}}^{v_i}, \theta_{\text{lati}}^{v_i}))$ of the vehicle $v_i$. Then a centroid $\mu_j$ will be selected, and based on that $\mu_j$, we can get the mapped division $\chi_i$.

Consequently, a horizontal window $(\chi_{i}-\Psi, \chi_{i}+\Psi)$ will be selected to get the finalized semantic mask. Within this window, if there is only one bounding box $\xi_{v}^{bb}$, it will be chosen as the target vehicle $v_i$, with all others being disregarded. Conversely, if multiple bounding boxes are detected within the window $(\chi_{i}-\Psi, \chi_{i}+\Psi)$, the one with the largest area contained within the window is chosen as the target vehicle $v_i$.
The mathematical representation for selecting $\chi_i$ based on the vehicle’s location and the mapped division from the BS's KB is as follows:
\vspace{-2mm}
\begin{equation}
\chi_i = \mathcal{K}(\mu_j(G^{v_i}(\Phi_{\text{long}}^{v_i}, \Phi_{\text{lati}}^{v_i}))),
\vspace{-2mm}
\end{equation}
where $\mathcal{K}(\mu_j(\cdot))$ denotes the mapping from a vehicle's geographical location to the corresponding division $\chi_i$ through centroid $\mu_j$.
The decision process for selecting the target vehicle $v_i$ from the identified bounding boxes within the window is as follows:
\begin{equation}
\vspace{-2mm}
\begin{split}
v_i = 
\begin{cases} 
\xi_{v}^{bb}, & \text{if } N(\xi_{v}^{bb}) = 1, \\
\arg\max \limits_{\xi_{v}^{bb}} \Gamma(\xi_{v}^{bb}), & N(\xi_{v}^{bb}) > 1.
\end{cases}
\end{split}
\label{eq:vehicleSelection}
\vspace{-2mm}
\end{equation}
Here $N(\xi_{v}^{bb})$ represents the number of bounding boxes inside the window $(\chi_{i}-\Psi, \chi_{i}+\Psi)$. The function $\Gamma(\xi_{v}^{bb})$ calculates the area of the bounding boxes inside the window $(\chi_{i}-\Psi, \chi_{i}+\Psi)$, ensuring that the most significant vehicle is selected within the window of the semantic mask.
\begin{figure}[t]
\centerline{\includegraphics[width=9cm]{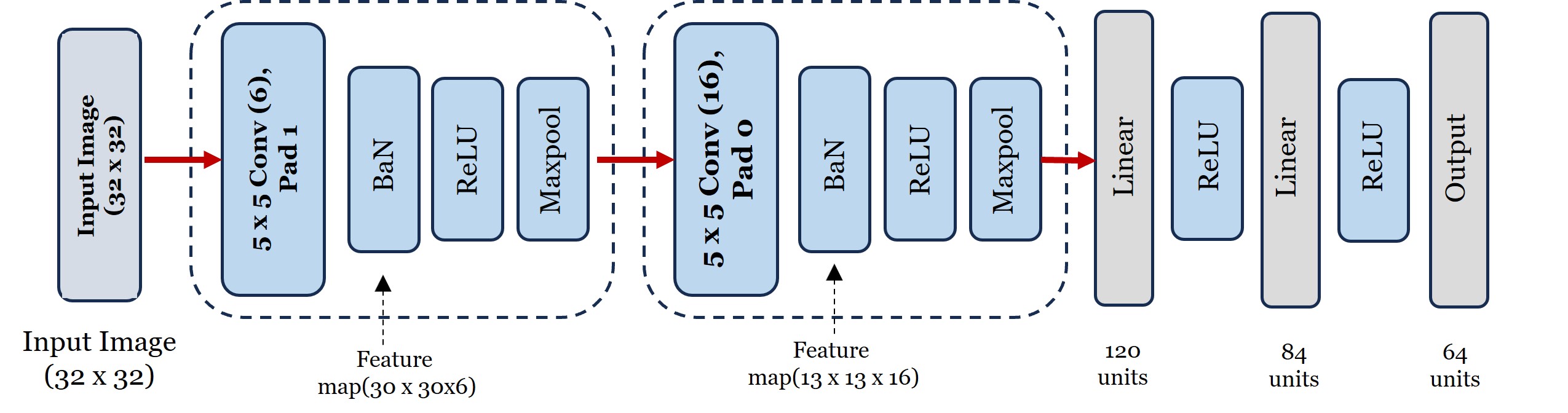}}
\vspace{-3mm}
\caption{LeNet 5 architecture.}
\label{Lenet}
\vspace{-5mm}
\end{figure}
\vspace{-2mm}

\begin{figure}[t]
\centerline{\includegraphics[width=9cm]{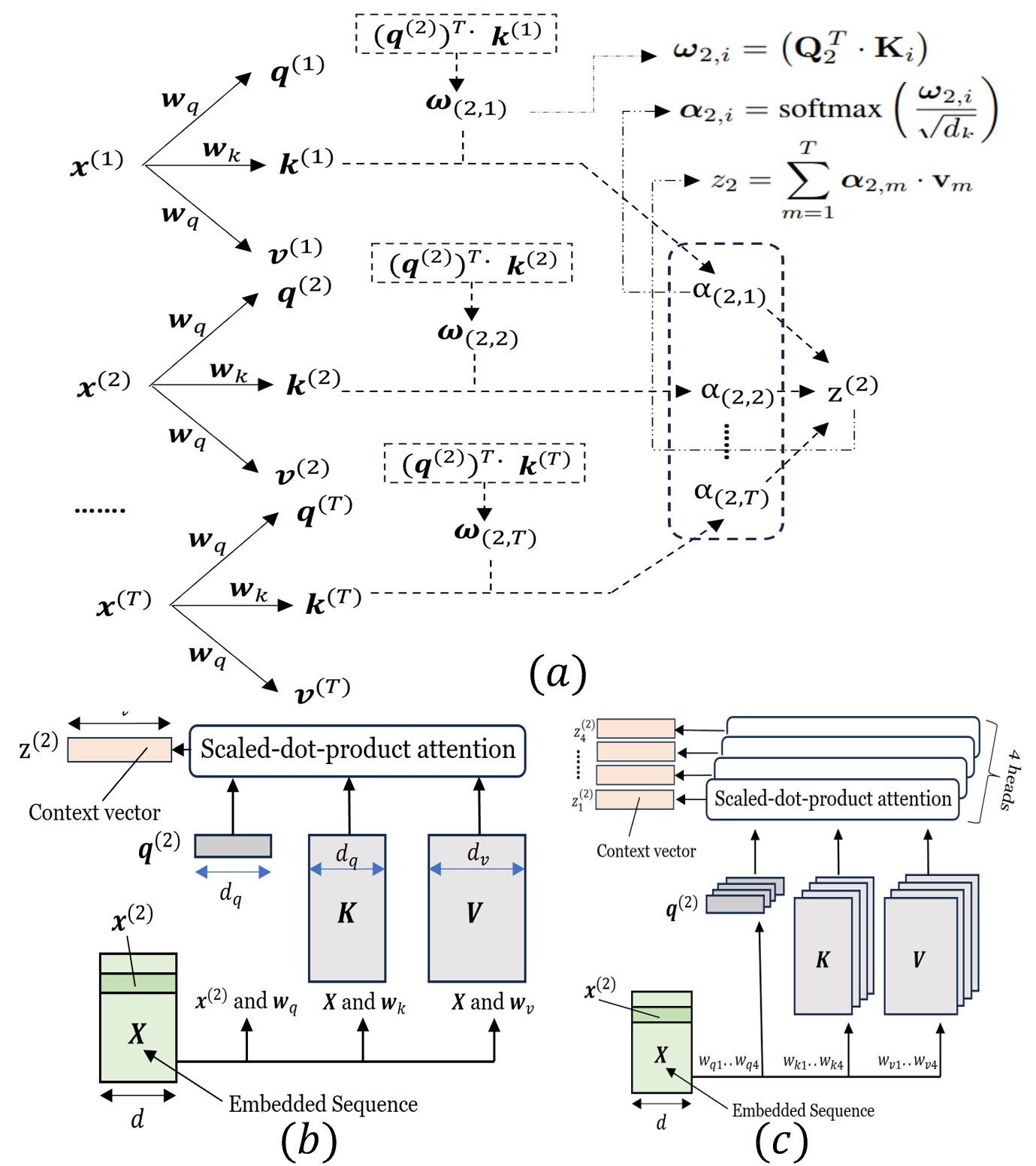}}
\vspace{-4mm}
\caption{(a) The overall procedures for computing the single-head attention
scores; (b) illustration of the single-head attention; (c) demonstration of the
multi-head attention \cite{adhikary2024holographic}.}
\vspace{-6mm}
\label{transformer}
\end{figure}
\subsection{Optimized Beam Selection Process}\label{BeamSelection}
\vspace{-2mm}
Our beam selection framework integrates two lightweight deep learning models. The first model processes binary semantic masks, while the second incorporates GPS data and the previous optimal beam index. The predictions of both models are then combined through a weighted entropy mechanism to dynamically select the final beamforming vector.
\subsubsection{Semantic Mask-Based Model}
Accurately predicting optimal beamforming vectors using binary semantic masks requires a model with high precision and computational efficiency. The LeNet5 model \cite{LeNet5} excels in this role due to its compact architecture and proficiency in classifying binary images, making it particularly suitable for real-time processing applications without compromising accuracy and performance. Additionally, its minimalistic design, which requires fewer computational resources, makes it highly suitable for real-time deployment in communication systems. LeNet5's combination of convolutional, subsampling, and FC layers facilitates efficient feature extraction from semantic information $\alpha_{t,\mathcal{E}}^{v_i}$, enabling precise selection of beamforming vectors from codebook $\mathbf{X}$ that contribute to enhanced communication link reliability for maintaining QoS. The architecture of the LeNet5 model is given in Fig. \ref{Lenet}. 


\subsubsection{Transformer Based Beam Selection Model}
In order to enhance the prediction of optimal beam indexes and ensure QoS across diverse environmental conditions, we integrate a transformer-based architecture into the framework. The transformer architecture is proficient in processing sparse inputs. In this study, we have designed a lightweight transformer architecture to predict the optimal beam index. This lightweight architecture exploits GPS positioning and the previous beam indices to maintain performance. This is particularly useful in scenarios where the semantic localization model struggles to identify vehicle information (e.g., semantic masks) in challenging environmental conditions. We have chosen self-attention for the transformer because of its computation efficiency across layers, ability to handle distant data correlations, and reduced sequence of operations needed. Transformers employ dot-product attention, which helps integrate highly efficient matrix multiplication algorithms, ensuring faster processing times and more efficient memory usage. The foundational operation of the transformer model is the categorization and processing of input vectors into queries (\(Q\)), keys (\(K\)), and values (\(V\)). These elements are combined to compute attention scores, facilitating an adaptive focus on pertinent information crucial for effective beamforming. The attention score can be calculated as follows \cite{adhikary2024holographic, vaswani2017attention}:
\vspace{-2mm}
\begin{equation}
\text{Attention}(\mathbf{Q}, \mathbf{K}, \mathbf{V}) = \text{softmax}\left(\frac{\mathbf{QK}^T}{\sqrt{d_k}}\right)\mathbf{V},
\vspace{-1mm}
\end{equation}
where \(d_k\) represents the dimensionality of the keys and queries. This formulation highlights the transformer's ability to prioritize different input data segments, improving the model's focus on important information for beam selection. Further elaboration on the self-attention mechanism is provided in Fig. (\ref{transformer}$a$), which illustrates the computation of attention scores to produce the context vector $y$ for an input sequence $X = \{x_1, x_2, ..., x_T\}$ of length $T$. These attention scores, $\alpha$, are normalized weights calculated from the softmax function applied to the raw attention weights $\omega$. The transformation of the input sequence into the queries, keys, and values is facilitated by the weight matrices $\mathbf{w}^q$, $\mathbf{w}^k$, and $\mathbf{w}^v$, each tailored to the specific dimensions $d_k$ and $d_v$, where $d_v$ is the dimension of the values. 

The architecture incorporates a single attention head that processes the input sequence within the scaled dot-product attention framework as demonstrated in Fig. (\ref{transformer}$b$). However, to enhance the model's capability to capture different aspects of the input data, multi-head attention is employed. This technique utilizes $d$ distinct attention heads, each equipped with its projection matrices $\mathbf{w}^Q_j$, $\mathbf{w}^K_j$, and $\mathbf{w}^V_j$ for queries, keys, and values respectively, as shown in Fig. (\ref{transformer}$c$). The output from the multi-head attention mechanism is aggregated as follows\cite{adhikary2024holographic, vaswani2017attention}:
\vspace{-2mm}
\begin{equation}
\text{MultiHead}(\mathbf{Q}, \mathbf{K}, \mathbf{V}) = \text{Concat}(\text{head}_1, ..., \text{head}_d)W^O,
\vspace{-1mm}
\end{equation}
with each head $j$ applying the attention function to project the input sequence through different subspaces. 
Additionally, the transformer incorporates a feed-forward network within its architecture, consisting of two linear transformations with an intervening ReLU activation function, represented by\cite{adhikary2024holographic, vaswani2017attention}:
\vspace{-2mm}
\begin{equation}
\text{FFN}(\mathbf{Y}) = \max(0, \mathbf{Y}W_1 + b_1)W_2 + b_2,
\vspace{-2mm}
\end{equation}
where $W_1$, $W_2$, $b_1$, and $b_2$ are the transformation parameters. Despite the variability in parameters across different layers, the linear transformation strategy remains uniform, ensuring consistency in processing across all positions within the sequence. 

\begin{algorithm}[!t]
\small
	\caption{Algorithm for the proposed solution approach for the BS}
	\label{alg1}
	\begin{algorithmic}[1]
		\renewcommand{\algorithmicrequire}{\textbf{Input: }}
		\renewcommand{\algorithmicensure}{\textbf{Output:}}
		\REQUIRE Captured image and GPS pair $(I(t, \mathcal{E}), \{G^{v}(\Phi_{\text{long}}^{v}, \Phi_{\text{lati}}^{v})\})$.
		\ENSURE  \textbf{optimal parameters for} transformer model: $\theta_t^*$, semantic model: $\theta_s^*$ , Optimal beam index $i^*_{x_i^L}$
        \\ \textbf{Initialization}: Neural Network Parameters $\theta_{s}$, $\theta_{t}$, $\textbf{S}_t$, $\textbf{S}_l$.
        \STATE Divide $\mathcal{I}$ into $\mathcal{I}_{\textrm{train}}$ and $\mathcal{I}_{\textrm{test}}$
		\FOR{$\forall I_i \in \mathcal{I}_{\textrm{train}} $}
        \STATE \textbf{Calculate:} $\alpha_{t,E}^\mathcal{V}$ using (\ref{eq14}).
        \STATE  \textbf{Evaluate constraints:}(\ref{Opt_1_4:const5}), (\ref{Opt_1_4:const10})
        \STATE \textbf{Calculate:} binary mask using (\ref{eq20}).
        \STATE  \textbf{Evaluate constraints:}(\ref{Opt_1_4:const5}), (\ref{Opt_1_4:const6}), (\ref{Opt_1_4:const7}).
        \STATE \textbf{Calculate:} $\chi_{v_i}$ using the GPS $\{G^{v_i}(\Phi_{\text{long}}^{v_i}, \Phi_{\text{lati}}^{v_i})\})$.
        \STATE  \textbf{Evaluate constraints:}(\ref{Opt_1_4:const8}), (\ref{Opt_1_4:const9}).
        \STATE \textbf{Target vehicle:} Extract target vehicle information $\alpha_{t,E}^{v_i}$ using window $(\chi_{v_i}-\Psi, \chi_{v_i}+\Psi)$.
		\STATE \textbf{Store:} ($\alpha_{t,E}^{v_i}$, $i_{x_i^L}$) to $\textbf{S}_s$.
            \STATE \textbf{Store:} (\{$G^{v_i}(\Phi_{\text{long}}^{v_i}, \Phi_{\text{lati}}^{v_i})$, ${(i-1)}_{x_(i-1)^L}$ \}, $i_{x_i^L}$) to $\textbf{S}_t$.
        \ENDFOR
        \STATE \textbf{Train:} LeNet5 model using the data stored in $\textbf{S}_l$.
        \WHILE{(Training==True)}
        \STATE Do the forward pass for the mini-batch $\mathcal{\textbf{S}^{\textrm{mini}}}_s$
        \STATE \textbf{Calculate:} Loss using eq (\ref{loss}).
        \STATE \textbf{Update:} $\theta_s$ by using the Adam optimizer.
        \ENDWHILE
        \STATE \textbf{store:} the optimal model $\theta_s^*$

        \STATE \textbf{Train:} Transformer model using the data stored in $\textbf{S}_s$.
        \WHILE{(Training==True)}
        \STATE Do the forward pass for the mini-batch $\mathcal{\textbf{S}^{\textrm{mini}}}_t$
        \STATE \textbf{Calculate:} Loss using (\ref{loss}).
        \STATE \textbf{Update:} $\theta_t$ by using the Adam optimizer \cite{attention_to_mPox}.
        \ENDWHILE
        \STATE \textbf{store:} the optimal model $\theta_s^*$.
	\WHILE{(Test==True)}
        \FOR{$\forall I_i \in \mathcal{I}_{\textrm{test}} $}
         \STATE \textbf{Calculate:} $\alpha_{t,E}^\mathcal{V}$ using (\ref{eq14}).
         \STATE  \textbf{Evaluate constraints:}(\ref{Opt_1_4:const5}), (\ref{Opt_1_4:const10})
        \STATE \textbf{Calculate:} binary mask using (\ref{eq20}).
        \STATE  \textbf{Evaluate constraints:}(\ref{Opt_1_4:const5}), (\ref{Opt_1_4:const6}), (\ref{Opt_1_4:const7}).
        \STATE \textbf{Calculate:} $\chi_{v_i}$ using the GPS $\{G^{v_i}(\Phi_{\text{long}}^{v_i}, \Phi_{\text{lati}}^{v_i})\})$.
        \STATE  \textbf{Evaluate constraints:}(\ref{Opt_1_4:const8}), (\ref{Opt_1_4:const9}).
        \STATE \textbf{Target vehicle:} Extract target vehicle information $\alpha_{t,E}^{v_i}$ using window $(\chi_{v_i}-\Psi, \chi_{v_i}+\Psi)$.
	  \STATE \textbf{Evaluate:} The optimal beam index $i_{\mathbf{x}_i^L}$ from the trained model parameters $\theta_s^*$.
        \STATE \textbf{Evaluate:} The optimal beam index $i_{\mathbf{x}_i'^L}$ from the trained model parameters $\theta_t^*$.
        \STATE  \textbf{Evaluate constraints:}(\ref{Opt_1_4:const1}), (\ref{Opt_1_4:const3}), (\ref{Opt_1_4:const11}).
        \STATE \textbf{select:} Optimal beam index using (\ref{best_beam}).
        \ENDFOR
        \ENDWHILE
	\end{algorithmic} 
\end{algorithm}
\vspace{-4mm}
\subsection{Integrated Beam Selection Strategy}
\begin{figure}[t]
\centerline{\includegraphics[width=9cm]{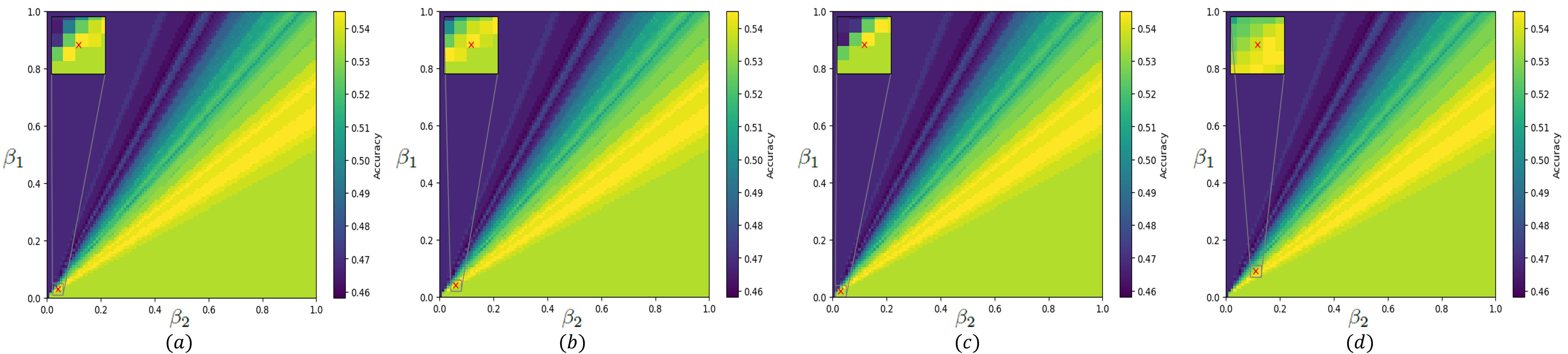}}
\vspace{-3mm}
\caption{Calibration of confidence weights \( \beta_1 \) and \( \beta_2 \) through grid search to maximize prediction accuracy: $(a)$ scenario 1; $(b)$ scenario 2; $(c)$ scenario 3; $(d)$ scenario 7.}
\label{beta}
\vspace{-5mm}
\end{figure}

We utilize a hybrid model for optimal beam selection that involves a semantic localization empowered model $s_1(\cdot)$ and a transformer-based model $s_2(\cdot)$ to find the optimal beamforming vector in diverse environmental scenarios. Both models' predictions are combined to make the final decision. The integration strategy considers the confidence levels of each model, which are based on the entropy of the model. Entropy measures the uncertainty or unpredictability in the model's predictions. A lower entropy value indicates that a model's predictions are more consistent and thus can be considered more reliable.
In contrast, a higher entropy signifies greater unpredictability, suggesting that the model is less specific about its predictions. Utilizing entropy allows us to objectively evaluate and compare the confidence levels of the two models' predictions. We use a dynamic decision-making criterion that takes into account the entropies $\alpha_1$ and $\alpha_2$ of the models:
\vspace{-2mm}
\begin{equation}
\omega = \begin{cases} 
1, & \text{if }\beta_1\cdot\alpha_1 < \beta_2\cdot\alpha_2, \\
0, & \text{otherwise},
\vspace{-1mm}
\end{cases}
\end{equation}
\vspace{-1mm}
\begin{equation}
\mathbf{x}_i^{*L} = \omega \cdot \mathbf{x}_i^L + (1- \omega) \cdot \mathbf{x}_i'^L, \label{best_beam}
\vspace{-1mm}
\end{equation}
where $\mathbf{x}_i^L$ and $\mathbf{x}_i'^L$ are the selected beams from models $s_1(\cdot)$ and $s_2(\cdot)$, respectively. Here, $\alpha_j$ represents the entropy of the $j^{th}$ model which can be represented as:
\vspace{-1mm}
\begin{equation}
\vspace{-2mm}
\alpha_j = -\sum_{i}^{N} p_{\mathbf{x}_i^L} \log(p_{\mathbf{x}_i^L}), \quad j \in \{1, 2\},
\vspace{-2mm}
\end{equation}
where $p_{\mathbf{x}_i^L}$ represents the probability associated with class $\mathbf{x}_i^L$. In the context of our hybrid beam selection approach, this represent the probability assigned by each model to beamforming vector $\mathbf{x}_i^L$ and $\log(p_{\mathbf{x}_i^L})$ represents the natural logarithm of the probability $p_{\mathbf{x}_i^L}$. We also use weights, represented by $\beta_j$, to balance the confidence levels and uncertainties of the models' predictions for different scenarios. The optimal calibration of these weights, $\beta_1$ and $\beta_2$, is achieved through the employment of a grid search technique over the range 0.01 to 1.00, utilizing the test dataset specific to each scenario. This methodology aims to identify the $\beta_1$ and $\beta_2$ values that maximize model accuracy. The resultant accuracy metrics, corresponding to diverse values of $\beta_1$ and $\beta_2$, are delineated in Fig. \ref{beta}. Examination of this figure  illustrates the optimal $\beta$ values tailored to each scenario. The solution approach is given in Algorithm \ref{alg1}.

\textbf{Complexity Analysis of Algorithm \ref{alg1}:} 
The complexity of Algorithm \ref{alg1} can be divided into two steps: the semantic localization process and the beam selection process. The semantic localization process starts with the YOLOv8n model and then applies a K-means clustering technique. The inference time complexity of the YOLOv8n model can be analyzed by considering the forward pass complexity. Let \( \mathcal{M} \) denote the number of layers, \( W_i \times H_i \) represent the dimensions of the input image \( i \), \( \zeta \) be the kernel size, and \( C_{\dag} \) and \( C_{\ddag} \) be the number of input and output channels, respectively. The complexity for each convolutional layer is \( O(\zeta \cdot W_i \cdot H_i \cdot C_{\dag} \cdot C_{\ddag}) \). Summing over all layers, the overall forward pass complexity for inference is \( O(\mathcal{M} \cdot \zeta \cdot W_i \cdot H_i \cdot C_{\dag} \cdot C_{\ddag}) \). The K-means clustering involves finding the division \( \chi_i \) from the set \( \digamma \). For each point \( p_i = (\Phi_{\text{long}}^{v_i}, \theta_{\text{lati}}^{v_i}) \), the algorithm calculates the distance to each of the \( \digamma \) centroids, resulting in a complexity of \( O(\digamma \times 2) \) where \( 2 \) is the dimension of each point. 

For the beam selection process, the complexity of the semantic model can be calculated similarly to the YOLOv8n model by considering the forward pass complexity, which can be written as \( O(\mathcal{M} \cdot \zeta \cdot W_i \cdot H_i \cdot C_{\dag} \cdot C_{\ddag}) \). For the transformer model, the input \( X \) with shape \( (T, f_d) \) consists of sequences of length \( T \) and dimension \( f_d \). The single-headed self-attention process involves two primary operations: linear transformation of \( X \) to generate the query \( Q \), key \( K \), and value \( V \) matrices, each with the shape \( (T, f_d) \). This is achieved through the post-multiplication of \( X \) with three learned matrices, each having dimensions \( (f_d, f_d) \), resulting in a computational complexity of \( O(T \times f_d^2) \). When calculating the layer output, the softmax function, applied row-wise, and the computation of \( QK^T \) incur a complexity of \( O(T^2 \times f_d) \). Subsequent multiplication with \( V \) also shares this complexity, making the total complexity of the layer \( O(T^2 \times f_d + T \times f_d^2) \). With the employment of \( d \) number of heads in multi-head attention, the dimensions for \( d_k \) and \( d_v \) are adjusted accordingly, reducing the dimension of each head, yet the overall computational complexity remains \( O(T^2 \times f_d + T \times f_d^2) \).
\begin{figure*}[htbp]
\centerline{\includegraphics[width=16cm]{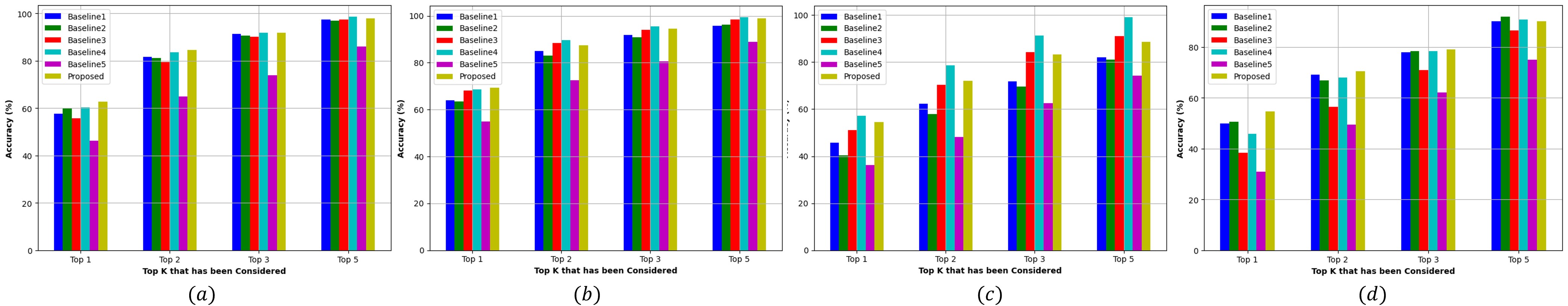}}
\vspace{-3mm}
\caption{Comparative analysis of baselines and proposed model performances across scenarios in top-K beam prediction accuracy: $(a)$ scenario 1; $(b)$ scenario 2; $(c)$ scenario 3; $(d)$ scenario 7.}
\vspace{-4mm}
\label{TopKaccuracy}
\end{figure*}
\vspace{-4mm}
\section{Performance Evaluation}\label{results}
\vspace{-2mm}
\begin{figure*}[htbp]
\centerline{\includegraphics[width=16cm]{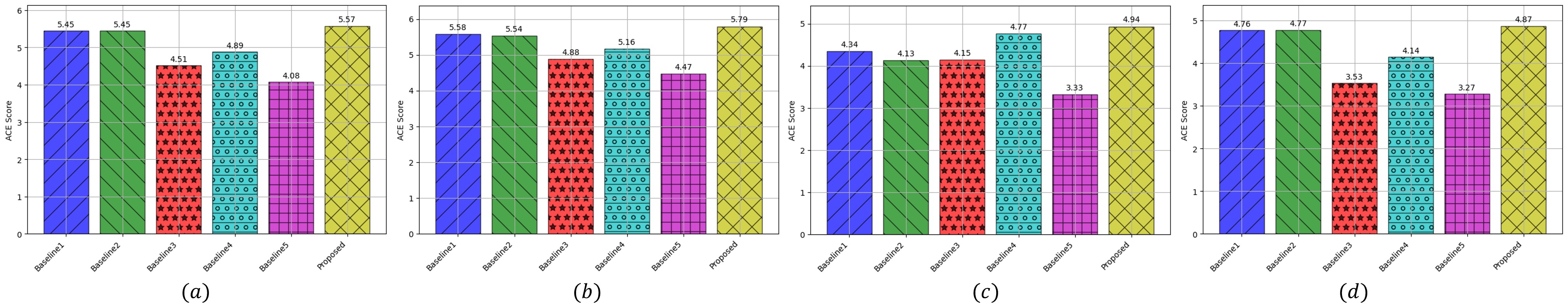}}
\vspace{-3mm}
\caption{ACE score comparison across different models: $(a)$ scenario 1; $(b)$ scenario 2; $(c)$ scenario 3; $(d)$ scenario 7.}
\vspace{-3mm}
\label{ACEScore}
\end{figure*}
In this section the performance of the proposed transformer and semantic localization empowered hybrid deep learning model is evaluated for the robust beamforming in mmWave communication. For the experimental settings, we consider a system featuring a BS equipped with a 16-element antenna array, operating in the 60 GHz frequency band. The RGB camera captures images at a resolution of 960x540. The target vehicle $v_i$ consists of a quasi-omni antenna at 60 GHz band. For simulation and result analysis, we used the DeepSense 6G dataset\cite{dataset}, which is a vast collection of real-world data designed to support advancements in deep learning research. This dataset is especially relevant for applications that require the integration of sensing, communication, and positioning technologies. It includes a diverse range of data types, such as mmWave wireless communication signals, GPS locations, RGB images, and LiDAR mappings. Each scenario in the dataset has detailed received power information across 64 transmit beams, which led to the creation of 64 distinct classes for our research. In our beam prediction approach, we used both GPS locations and RGB images from the dataset.
\vspace{-2mm}
\subsection{Experimental Setup}\label{AA}
For evaluating the performance of our method, we chose scenarios 1, 2, 3, and 7 from the DeepSense 6G dataset \cite{dataset}. Scenarios 1, 2, and 3 include multiple vehicles, with one being the target vehicle. To demonstrate how our method works for a single user, we used scenario 7.  We have compared our approach, with several baselines, described Table \ref{tab:table1}. The simulation parameters are listed in Table \ref{tab:tableParam}.
\renewcommand{\arraystretch}{1} 
\begin{table}[htbp]
  \vspace{-5mm}
  \centering
  \caption{Baseline Descriptions\label{tab:table1}}
  \vspace{-3mm}
  \resizebox{\columnwidth}{!}{
    \begin{tabular}{lll}
    \toprule
    \textbf{BaseLine} & \textbf{Semantic Ext.} & \textbf{Classifier} \\
    \midrule
    \textbf{1} & YOLOV8n & Lenet5 \\
    \textbf{2} & YOLOV8n & Lenet5 + GPS in FC \\
    \textbf{3} & N/A & Finetuned ResNet-152 + GPS in FC  \\
    \textbf{4} & N/A & Finetuned DenseNet-161 + GPS in FC  \\
    \textbf{5} & N/A & Finetuned VGG-16 + GPS in FC  \\
    \bottomrule
    \end{tabular}%
  }
  \vspace{-2mm}
\end{table}
To rigorously test the strength of our proposed methodology, we developed six distinct scenarios covering a broad spectrum of environmental and lighting conditions, leveraging image augmentation tools from \textit{imgaug} \cite{imgaug} and \textit{albumentations} \cite{albumentations}. These testing scenarios allow us to conduct an in-depth evaluation of the model's adaptability, strength, and capacity to generalize across varied conditions.
\begin{itemize}

    \item \textbf{Night-to-Day}: In this scenario, models are initially trained on nighttime images and subsequently evaluated on daytime images. We utilize scenario 3 from the DeepSense 6G dataset, selecting the last 700 images for training for their notably darker lighting conditions and the first 200 images for testing due to their significantly brighter lighting conditions.
    \item \textbf{Day-to-Night}: This scenario reverses the conditions of the previous one. Here, training is conducted using the first 700 images from scenario 3, while testing utilize the last 200 images. The models are trained on daytime images and evaluated against nighttime images to determine their ability to adapt to low-light environments.
    \item \textbf{Rainy Night Environment}: For the \textit{Rainy Night} environment scenario, models are trained on the same daytime images as those used in the \textit{Day-to-Night} scenario. However, for testing, they were faced with a set of the last 200 images from the previous scenario, now augmented with artificial rain, using the \textit{imgaug}\cite{imgaug} tool.
    \item \textbf{Snow-Covered Environments}: Employing scenario 1 from the dataset, models are trained under standard RGB images and evaluated under artificially snow-augmented images using the \textit{albumentations}\cite{albumentations} tool. We created four snowy environments (\textit{Snowy 01}, \textit{Snowy 02}, \textit{Snowy 03}, \textit{Snowy 04}) by varying the \textit{snow point} parameter from 0.3 to 0.9. An additional testing environment, \textit{Snowy 05}, is introduced by incorporating snowflakes via the \textit{imgaug}\cite{imgaug} tool, which creates an extreme environment that tests the models' reliability and adaptability by making it difficult for them to identify typical features.
    \item \textbf{Rainy Environments}:   To simulate varying intensities of rainy conditions, we utilize scenarios 1 and 2 from the dataset. Using the augmentation capabilities of the tool \textit{albumentations} for scenario 1, we methodically craft four distinct rainy environments. This is achieved by adjusting the \textit{brightness coefficient} and \textit{blur value} parameters within the ranges of 0.3 to 0.6 and 4 to 7, respectively. These adjustments result in the creation of four scenarios: \textit{Rainy 01}, \textit{Rainy 02}, \textit{Rainy 03}, and \textit{Rainy 04}, each originating from scenario 1. Further diversification of rainy conditions is accomplished through scenario 2, where we employ another augmentation tool, \textit{imgaug}. Here, we modify the \textit{speed} variable, setting its values between 0.1 and 0.3, to generate three additional and distinct scenarios: \textit{Rainy 05}, \textit{Rainy 06}, and \textit{Rainy 07}.
    \item \textbf{Foggy Environment}: For the \textit{Foggy} environment scenario, we utilized \textit{imgaug} to augment images from scenario 2, creating foggy conditions. 
\end{itemize}
\begin{figure}[t]
\centerline{\includegraphics[width=9cm]{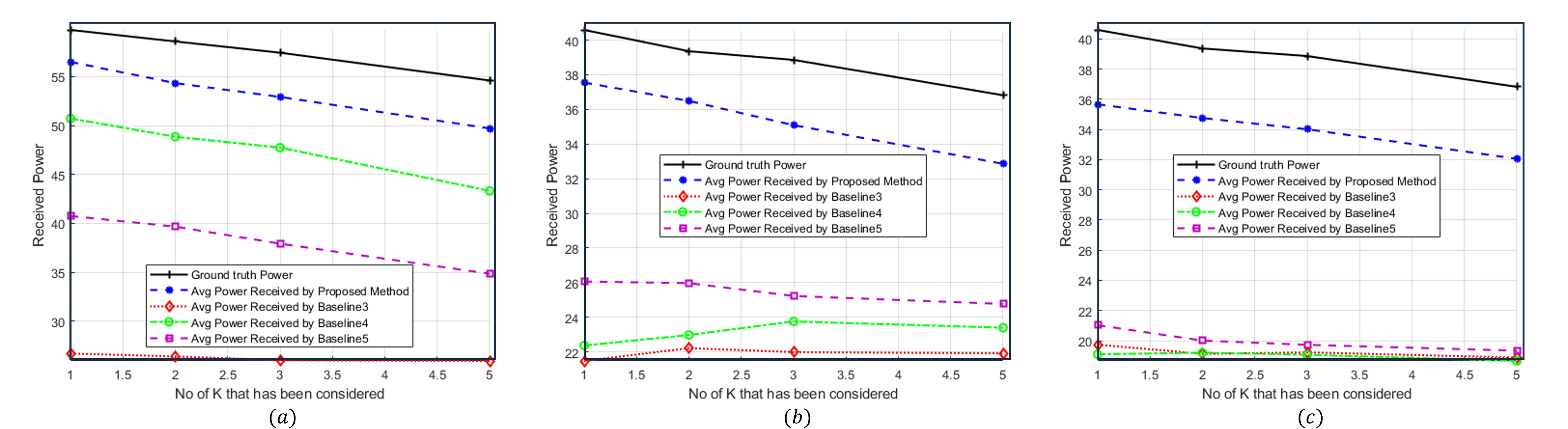}}
\vspace{-2mm}
\caption{Performance comparison among baseline 3, 4, 5 and proposed method in different lightning conditions: $(a)$ \textit{Night-to-Day}; $(b)$ \textit{Day-to-Night}; $(c)$ \textit{Rainy Night} environment.}
\vspace{-5mm}
\label{day_night_rainy_night}
\end{figure}

\renewcommand{\arraystretch}{1}
\begin{table}[htbp]
  \vspace{-2mm}
  \centering
  \caption{Simulation Parameters\label{tab:tableParam}}
  \vspace{-3mm}
    \begin{tabular}{ccc}
    \toprule
    \textbf{Parameters} & \textbf{LeNet5} & \textbf{Baselines} \\
    \midrule
    \textbf{Input Image Size} & 32 $\times$ 32 & 224$\times$ 224\\
    \textbf{Batch} & 32 & 32 \\
    \textbf{Epochs} & 30 & 30 \\
    \textbf{Learning rate}  & $10^-3$ & $10^-3$\\
    \textbf{LR Reduction Factor}  & 0.1 & 0.1 \\
    \textbf{Optimizer} & Adam & Adam  \\
    \textbf{Loss Function} & Cross Entropy & Cross Entropy \\
    \bottomrule
    \vspace{-5mm}
    \end{tabular}%
    \vspace{-3mm}
  \label{tab:parameters}%
\end{table}

\subsection{Numerical Results}\label{AA}
The comparison of various baselines across the scenarios 1,2,3 and 7 in terms of beam prediction accuracy is illustrated in Fig. \ref{TopKaccuracy}. From the figure, we can observe that the proposed model consistently surpasses the performance of baseline 1 and baseline 2. For instance, in scenario 1, the proposed model attains a Top-1 accuracy of 62.68\%, surpassing baseline 1 and baseline 2, which achieve 57.73\% and 59.79\% accuracy, respectively. Further scrutiny into scenario 2 reveals that the proposed model's Top-1 accuracy escalates to 69.40\%, significantly outperforming baseline 1 and baseline 2, which report accuracies of 63.87\% and 63.37\%, respectively. In scenario 3, despite the apparent challenges and the fluctuating dynamics of beam prediction, the proposed model upholds its performance with a Top-1 accuracy of 54.56\%, outstripping baseline 1 and baseline 2. The comprehensive analysis culminates with scenario 7, wherein the proposed model, with a Top-1 accuracy of 54.65\%, continues to exhibit its dominance over baseline 1 and baseline 2, which achieve 50.00\% and 50.58\%, respectively.
Baseline 1 employs semantic mask $\alpha_{t,E}^\mathcal{V}$, for encompassing all vehicles, and baseline 2 enhances this approach by integrating GPS information of the target vehicle into its fully connected layer. In contrast, in the proposed method we localized the target vehicle $v_i$ in the semantic mask, $\alpha_{t,E}^{v_i}$. By effectively localizing the target user $v_i$, the proposed model achieves better accuracy thereby maximizes the data-rate of the user vehicle $v_i$.
However, baselines 3 and 4, which do not utilize target vehicle localization, still attain performance levels comparable to that of the proposed method.

This is primarily because these baselines utilize RGB images for training, which allows them to recognize and remember the target vehicle by its distinctive shape and color. Since the dataset features the same target vehicle within each scenario, these methods can effectively capture its characteristics. This approach raises concerns regarding its real-world applicability. The dataset's controlled conditions, with the repetition of the same vehicle, do not reflect the diversity and variability encountered in real-world data. In a more dynamic setting, where the array of vehicle shapes and colors is vast and varied, the performance of baseline 3 and baseline 4 is expected to degrade as it can not localize the target vehicle $v_i$ effectively. This potential drop in effectiveness outside of the dataset's controlled environment highlights a critical limitation, questioning the robustness of these methods when faced with the complexities of real-world scenarios. 
Additionally, the parameter counts of baseline 3 and 4 present a significant consideration. These models possess a greater number of parameters compared to the proposed model, potentially impacting QoS due to the extended computational time required by more complex models. In high mobility applications, timely beam prediction is crucial to maintaining optimal performance. The increased computation time associated with these larger models can lead to significant delays in predicting the optimal beams. As a result, the system might have to rely on sub-optimal beams during this delay period, leading to reduced QoS. Therefore, to maintain performance, we require models that are not only lightweight but also efficient in predicting the optimal beams.  This observation underscores the drawbacks of traditional performance metrics, which primarily assess accuracy while neglecting to consider the computational demands caused by the complexity of the model. This has led to the development of a new metric named ACE, which can be represented as follows:
\begin{figure}[t]
\centerline{\includegraphics[width=9cm]{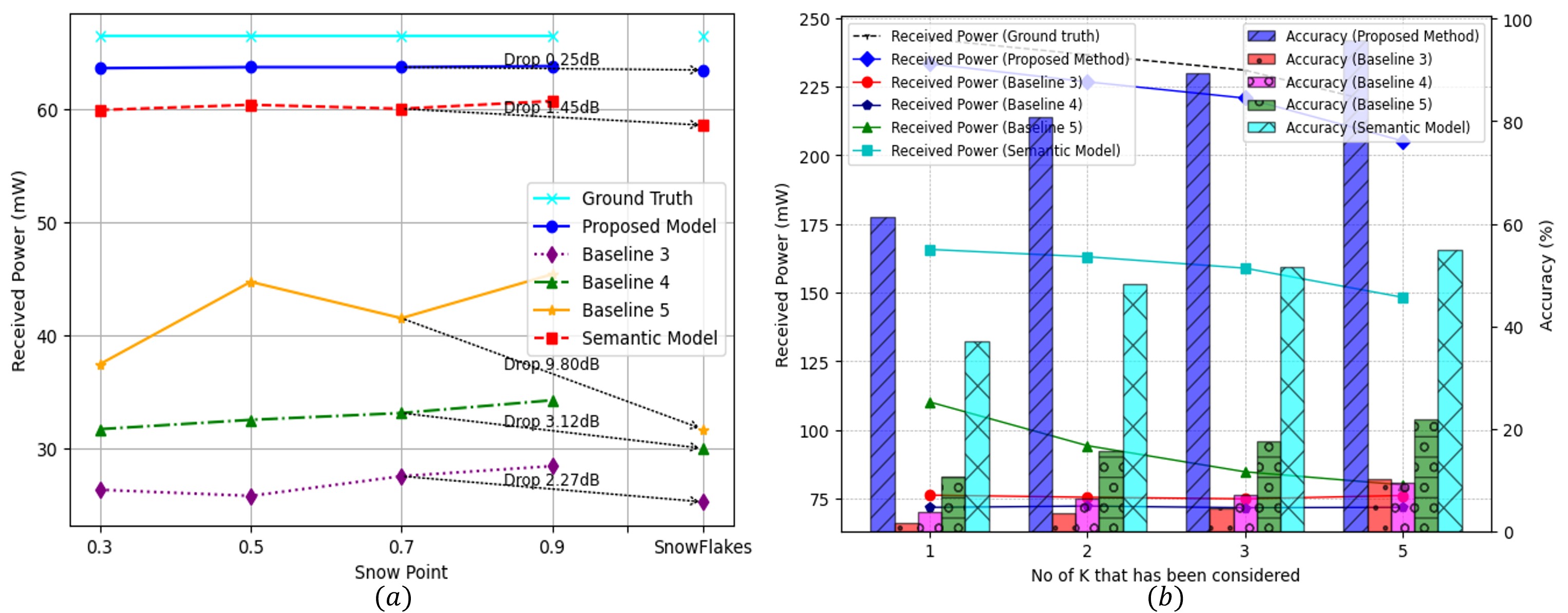}}
\vspace{-3mm}
\caption{Performance comparison of different baselines and proposed model in snowy and foggy environments.}
\vspace{-3mm}
\label{Snowy}
\end{figure}
\vspace{-3mm}
\begin{equation}
    \Phi_{\textrm{ACE}} = \frac{\Delta}{\log(1 + \Pi)}.
    \vspace{-2mm}
\end{equation}

Herein, \(\Delta\) represents the averages of Top-k accuracies achieved by the model, while \(\Pi\) denotes the number of parameters utilized by the model. 
The logarithmic scaling, \(\log(1 + \Pi)\), adjusts for the principle of diminishing returns associated with the augmentation of parameters in already large models. 
\begin{figure}[t]
\centerline{\includegraphics[width=9cm]{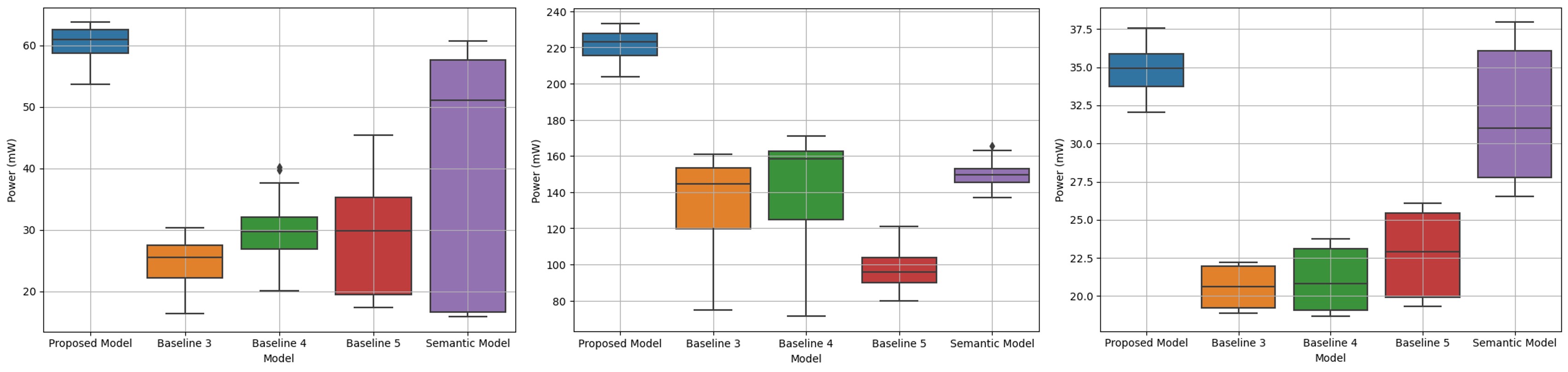}}
\vspace{-3mm}
\caption{Box plots of average received power of top-k beams across various methods and environments: $(a)$scenario 1; $(b)$scenario 2; $(c)$scenario 3.}
\vspace{-7mm}
\label{Distribution}
\end{figure}

\begin{table}[t]
\centering
\vspace{-3mm}
\caption{Top-K accuracy achieved in rainy scenarios}
\vspace{-3mm}
\label{table_rainy}
\resizebox{\columnwidth}{!}{%
\begin{tabular}{lccccccccl}
\toprule
 & \textbf{Rainy 01} & \textbf{Rainy 02} & \textbf{Rainy 03} & \textbf{Rainy 04} & \textbf{Rainy 05} & \textbf{Rainy 06} & \textbf{Rainy 07} & & \textbf{Method} \\
\midrule
Top 1 & 1.44 & 2.47 & 7.01 & 13.19 & 34.11 & 33.11 & 33.44 & & \\
Top 2 & 3.29 & 3.71 & 11.13 & 22.47 & 44.81 & 44.48 & 45.81 & & Baseline 3 \\
Top 3 & 4.74 & 5.97 & 15.25 & 28.04 & 51.33 & 51.67 & 52.34 & & \\
Top 5 & 7.42 & 9.89 & 21.64 & 38.55 & 58.36 & 59.69 & 59.36 & & \\
\midrule
Top 1 & 2.68 & 7.01 & 16.28 & 24.53 & 35.11 & 38.96 & 38.46 & & \\
Top 2 & 5.97 & 13.6 & 27.21 & 38.76 & 49.33 & 50.66 & 51.57 & & Baseline 4 \\
Top 3 & 10.51 & 17.52 & 33.6 & 47.01 & 56.68 & 56.85 & 57.52 & & \\
Top 5 & 15.05 & 23.5 & 45.15 & 61.64 & 62.54 & 63.21 & 63.54 & & \\
\midrule
Top 1 & 3.91 & 5.15 & 4.32 & 6.18 & 10.2 & 13.87 & 8.02 & & \\
Top 2 & 6.18 & 6.39 & 6.39 & 9.27 & 16.55 & 19.06 & 13.54 & & Baseline 5 \\
Top 3 & 7.21 & 8.69 & 8.86 & 10.1 & 20.73 & 23.74 & 17.22 & & \\
Top 5 & 11.13 & 13.81 & 13.6 & 15.46 & 27.92 & 30.1 & 21.07 & & \\
\midrule
Top 1 & 1.03 & 1.03 & 1.03 & 1.03 & 30.36 & 29.59 & 28.92 & & \\
Top 2 & 3.09 & 3.09 & 3.09 & 3.09 & 41.63 & 41.80 & 38.96 & & Semantic Model \\
Top 3 & 5.77 & 5.77 & 5.77 & 5.77 & 44.98 & 45.65 & 42.14 & & \\
Top 5 & 8.88 & 8.88 & 8.88 & 8.88 & 49.49 & 50.56 & 44.98 & & \\
\midrule
Top 1 & \textbf{54.85} & \textbf{54.85} & \textbf{54.85} & \textbf{54.85} & \textbf{59.7} & \textbf{58.86} & \textbf{61.71} & & \\
Top 2 & \textbf{77.53} & \textbf{77.53} & \textbf{77.53} & \textbf{77.53} & \textbf{79.43} & \textbf{79.26} & \textbf{80.77} & & Proposed Model\\
Top 3 & \textbf{87.63} & \textbf{87.63} & \textbf{87.63} & \textbf{87.63} & \textbf{88.13} & \textbf{88.63} & \textbf{89.46} & & \\
Top 5 & \textbf{98.14} & \textbf{98.14} & \textbf{98.14} & \textbf{98.14} & \textbf{94.65} & \textbf{94.98} & \textbf{95.48} & & \\
\bottomrule
\vspace{-5mm}
\end{tabular}
\vspace{-3mm}
}
\end{table}
This mathematical formulation promotes the development of models that are not only accurate but also efficient in their use of parameters. Models that manage to achieve high accuracy with fewer parameters are deemed more efficient and are thus favored by this metric. This encourages the design of models that strike a delicate balance between accuracy and model size, which is particularly important in scenarios where faster computation is required. Therefore, the ACE measure becomes an important standard, supporting models that combine high accuracy with a small number of parameters.
The ACE score achieved by the baselines are shown in Fig. \ref{ACEScore}. From the figure it can be seen that the proposed model achieves the highest ACE score as it offers standard performance despite being a lightweight model.

Figs. \ref{day_night_rainy_night}$(a)$, \ref{day_night_rainy_night}$(b)$, and \ref{day_night_rainy_night}$(c)$ present the comparative performance of the proposed approach against established baselines in \textit{Day-to-Night}, \textit{Night-to-Day}, and \textit{Rainy Night} settings, respectively. Through these comparisons, it becomes evident that our method maintains superior performance margins across all tested scenarios.
In the \textit{Night-to-Day} environment depicted in Fig. \ref{day_night_rainy_night}$(a)$, our analysis compares the average received power across baselines for varying top-K selected beams. The proposed method shows a modest reduction in received power by 5.54\% compared to the ground truth when we select the the top-1(optimal) beam , demonstrating its reliability. In contrast, alternative baselines (baseline 3, baseline 4, and baseline 5) exhibit significantly larger decreases in received power: 55.34\%, 15.14\%, and 31.82\% for the top-1 beam, indicating a pronounced sensitivity to lightning changes. This pattern of performance degradation remains consistent across top 2, 3, and 5 beam selections, underscoring the robustness of our proposed methodology. Extending this analytical framework to scenarios shown in Figs. \ref{day_night_rainy_night}$(b)$ and \ref{day_night_rainy_night}$(c)$, we note similar performance trends. 

Fig. \ref{Snowy} illustrates the comparative performance between the proposed hybrid beam selection model and various baselines in the \textit{Snowy} and \textit{Foggy} environments. Fig. \ref{Snowy}$(a)$ compares the performance of the proposed model with several baseline models, including baseline 3, baseline 4, baseline 5, and the standalone semantic model $s_1(\cdot)$. The x-axis represents the intensity of snow in the testing set. For the last data point on the x-axis, we added snowflakes to simulate a scenario similar to \textit{Snowy 03}, with the snow intensity of the test image set maintained at 0.7. The y-axis displays the received power for the top-1 (optimal) predicted beam by each model. The figure shows that the proposed model consistently achieved optimal performance from \textit{Snowy 01} to \textit{Snowy 05}. Notably, in \textit{Snowy 05}, the received power slightly dropped to 63.43 mW from 63.68 mW (achieved in \textit{Snowy 03}), a minimal decrease of only 0.25 mW upon adding snowflakes to the scenario. This minimal performance drop underscores the robustness of the model in adverse conditions. Conversely, the semantic model $s_1(\cdot)$, while adaptable, saw a more significant decrease in received power to 58.56 mW from 60.01 mW (achieved in \textit{Snowy 03}) in the final scenario. On the contrary, baseline 4 and baseline 3 dropped received power to 30.04 mW and 25.32 mW from 33.16 mW and 27.59 mW, respectively. Baseline 5 also exhibited a significant decline to 31.73 mW from 41.53 mW, emphasizing the difficulty of maintaining performance under heavy snow conditions.

Fig. \ref{Snowy}$(b)$ compares the performance between the proposed method and other baselines in \textit{Foggy} weather. As we can see, the proposed method performs effectively in foggy conditions, consistently maintaining received power close to the ground truth. Specifically, the decrease in received power for the proposed method is relatively low, standing at 3.69\% for the top-1 beam selection compared to the ground truth. This minor decrease showcases the method's robustness, even under challenging visibility conditions. Baseline 3, baseline 4, and baseline 5, on the other hand, show considerable drops in performance, emphasizing their vulnerability to foggy conditions. For the top-1 beam selection, baseline 3 and baseline 4 show decreases in received power by 68.57\% and 69.68\%, respectively, while baseline 5 shows a decrease of 54.57\%. These significant reductions point out the limitations of these models in adverse weather conditions. The semantic model performs better than baseline 3, baseline 4, and baseline 5 but still follows behind the proposed method. It provides a medium ground in performance, better mitigating the effects of the foggy environment than the other baselines with a loss in received power of 31.63\% for the top-1 beam selection, but less effectively than the proposed approach.

The performance of the baselines and proposed method in the rainy environments are given in Table \ref{table_rainy}. From the table, we can see the performance of the models in terms of Top-k accuracy of beam prediction. As discussed earlier \textit{Rainy 01}, \textit{Rainy 02}, \textit{Rainy 03} and \textit{Rainy 04} are designed using \textit{albumentation}\cite{albumentations} from scenario 1. For these four scenarios, the difficulty level has been increased incrementally. From the table, we can see that the performance has increased to a certain level for baselines 3, 4, and 5 for these scenarios. However, the proposed model maintained similar performance across all four scenarios. Also, the semantic model $s_1(\cdot)$ constantly maintains the same poor performance. That is because, for the semantic model, the YOLOv8n cannot detect any possible vehicle from these scenarios. For the proposed model, as the semantic model does not detect any vehicle, only the transformer model predicts the beams using the GPS information. Still, the proposed model can outperform other baselines. This result shows the effectiveness of the transformer model's integration in the proposed solution architecture. For \textit{Rainy 05}, \textit{Rainy 06}, and \textit{Rainy 07}, we used the \textit{imgaug}\cite{imgaug} tool for adding rains to the test set. From the table, it can be seen that in these scenarios, the proposed model outperforms others as it leverages GPS and semantic information.

In Fig. \ref{Distribution}, the box plots show how the average received power for top-K beams varies for different baselines in different environments. Fig. \ref{Distribution}$(a)$ shows the box plot of average received powers for top-K beams in scenario 1. We can observe that the proposed method's performance has less variance in the received power while keeping the received power maximum. The other baselines showed huge variance with lower received power. The standalone semantic model $s1(\cdot)$ showed good performance for some environments; however, in some situations, it shows poor performance, which describes the importance of the integration of transformer architecture in the proposed model. Fig. \ref{Distribution}$(b)$ and \ref{Distribution}$(c)$ showed the box plot of average received powers for top-K beams in scenarios 2 and 3, respectively. We can also see similar trends here that prove the effectiveness of the proposed model. The proposed model achieves an increase in average received power ranging from 22.83\% to 42.10\% compared to baseline methods in dynamic environments, considering the ground truth power values.

\begin{figure}[t]
\centerline{\includegraphics[width=7.5cm]{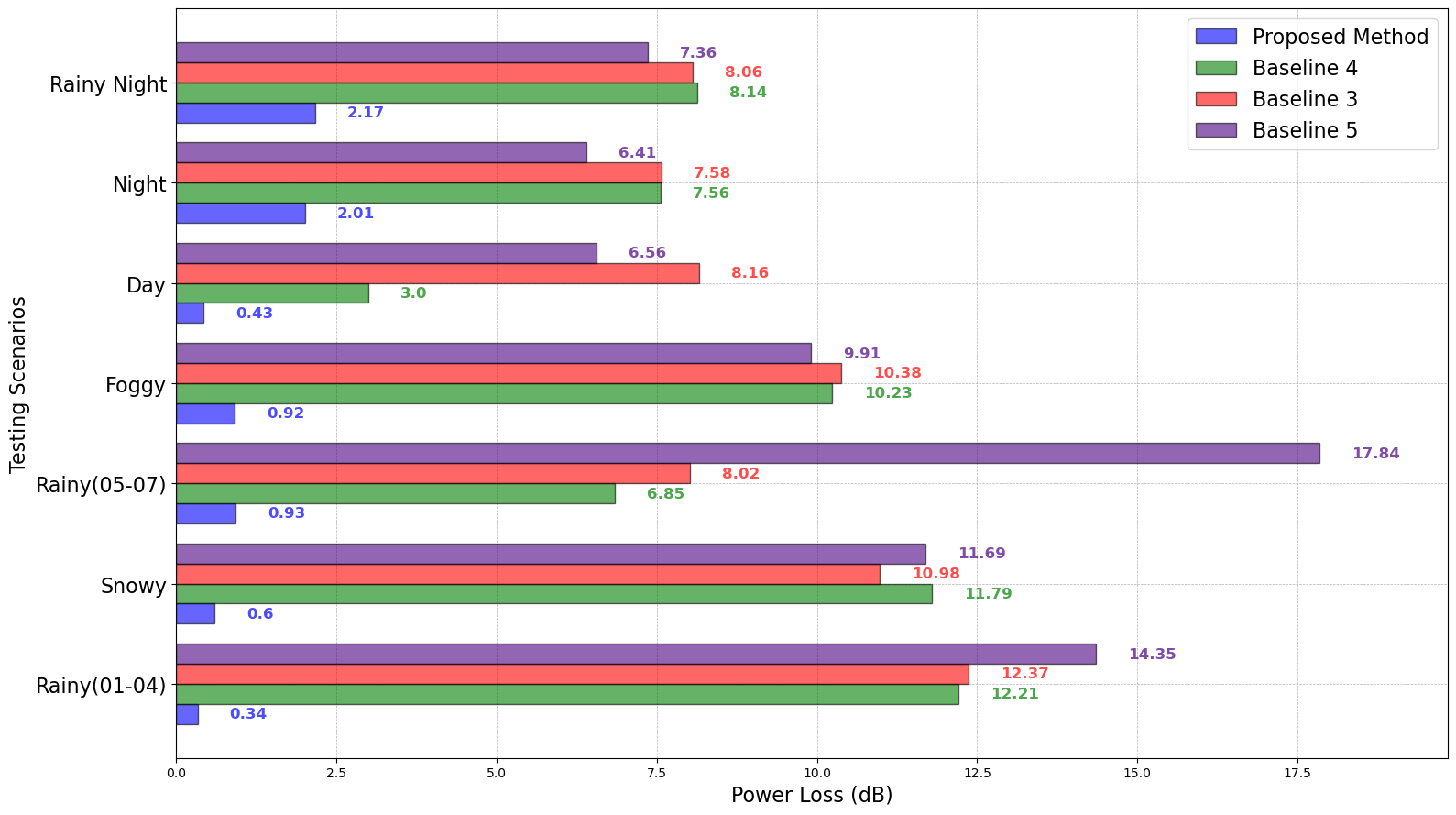}}
\vspace{-3mm}
\caption{Comparison of average power loss among baseline 3, 4, 5 and proposed method.}
\vspace{-4mm}
\label{fig10}
\end{figure}
For comparing the performance of the proposed model with a more robust metric we used the average power loss as proposed in \cite{position}. The average power loss can be defined as\cite{position}, 
\vspace{-2mm}
\begin{equation}
\mathcal{P}_{loss[DB]} = \frac{1}{C}\sum_{c=1}^{C}\frac{P_{\mathbf{b}^*}^c-P_{x}}{P_{\hat{\mathbf{b}}}^c-P_{x}}
\vspace{-2mm}
\end{equation}
Here, $P_{\mathbf{b}^*}^c$ and $P_{\hat{\mathbf{b}}}^c$ represent the power of the ground truth beam and predicted beam in sample $c$ respectively. $P_{x}$ is the noise power, the average smallest power per sample. In Fig. \ref{fig10}, we present a comparative analysis of the power loss between our proposed method and the baselines across various environmental conditions. From the figure, we can observe that the average power loss varies from 0.34 dB to 2.17 dB for different environments, which is much lower compared to baselines 3,4 and 5. Baselines 3, 4, and 5 vary from 7.58 dB to 12.37 dB, 6.85 dB to 12.21 dB, and 6.41 dB to 17.84 dB, respectively. The results indicate that the power loss is significantly lower when employing the proposed method. 
This enhanced performance demonstrates the efficacy of our model in optimizing power consumption while maintaining high prediction accuracy. The lower power loss results in better energy efficiency. Thus, the proposed method offers a more robust and sustainable solution for power management in diverse environmental conditions. 
\section{Conclusion} \label{conclusion}
This paper presents a robust beamforming technique that ensures consistent QoS despite fluctuations in environmental conditions. An optimization problem has been formalized to maximize users' data rates. In order to solve the NP-hard optimization problem, we break it down into two subproblems: the semantic localization problem and the optimal beam selection problem. We introduced a novel method for semantic localization that leverages the strengths of K-means clustering and the YOLOv8n model. To address the beam selection problem, we proposed a hybrid architecture that harnesses information from various data sources for accurate beam prediction. Furthermore, we employed lightweight transformer and CNN models to ensure energy efficiency, QoS, and performance, combining their predictions using a weighted entropy-based approach. To assess the trade-off between accuracy and computational efficiency, a novel metric called ACE is introduced. The proposed model's robustness was evaluated through extensive simulations across six diverse scenarios. The results demonstrate that the proposed model achieves improvements in average received power ranging from 22.83\% to 42.10\%, ACE improvements from 0.10 to 1.61, and a decrease in average power loss ranging from 2.57 dB to 16.91 dB, compared to baseline methods across various environmental conditions.

\vspace{-3mm}
\bibliographystyle{IEEEtran}
\bibliography{references}
\end{document}